\documentclass[journal=jacsat,manuscript=article]{achemso}
\usepackage[version=3]{mhchem} % Formula subscripts using \ce{}
\usepackage{xcolor}
\usepackage{comment}
\usepackage{xr}
\usepackage{soul}
\usepackage{tikz}
\usetikzlibrary{positioning}
\usetikzlibrary{calc}
\usepackage[table]{xcolor} 
\usepackage{hyperref}
\usepackage{enumitem}
\usepackage{amssymb}
\usepackage{array}
\usepackage{amssymb}
\usepackage{longtable}

\SectionNumbersOn

\newcommand{\red}[1]{\textcolor{red}{#1}}
\newcommand{\blue}[1]{\textcolor{blue}{#1}}
\newcommand{\green}[1]{\textcolor{green}{#1}}

\title[An \textsf{achemso} demo]
  {Martini Mapper: An Automated Fragment-Based Mapping Algorithm for Developing Coarse-Grained Models within the Martini 3 Framework}

\author{Kevin V. Bigting}
\affiliation{Department of Computer Science and Engineering, Louisiana State University, Baton Rouge, LA, 70803}
\altaffiliation{K.V.B. and S.N. contributed equally to this work.}
\author{Shubhadeep Nag}
\affiliation{Department of Chemical Engineering, Louisiana State University, Baton Rouge, LA, 70803}
\altaffiliation{K.V.B. and S.N. contributed equally to this work.}
\author{Yaxin An}
\affiliation{Department of Chemical Engineering, Louisiana State University, Baton Rouge, LA, 70803}
\email{yxan@lsu.edu}

\begin{document}

\begin{tocentry}
\includegraphics{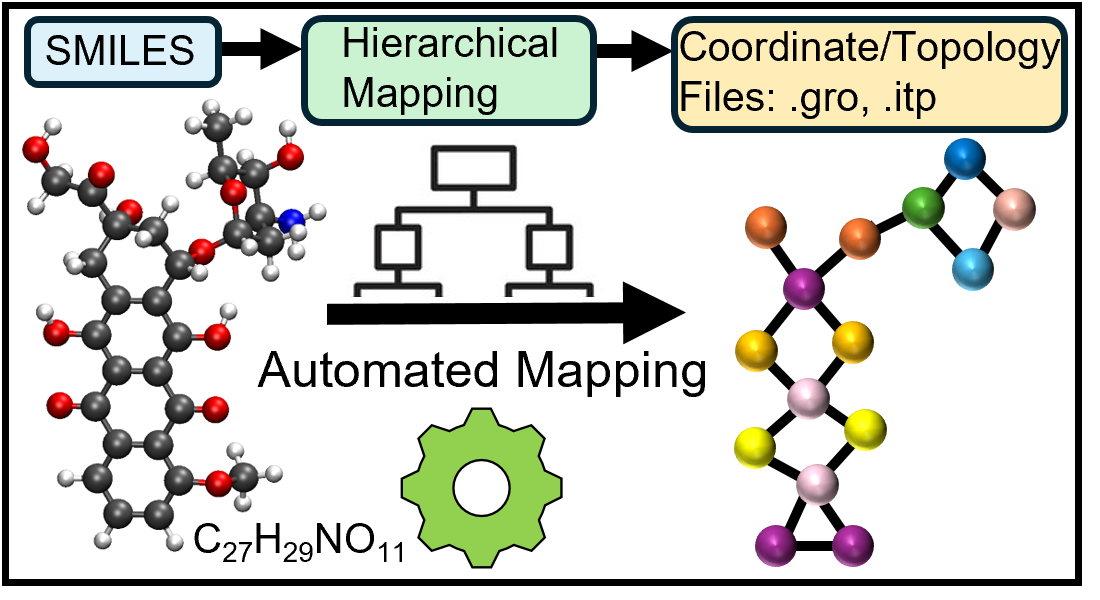}
\end{tocentry}

\begin{abstract}
Coarse-graining (CG) reduces molecular details to extend the time and length scales of molecular dynamics simulations to microseconds and micrometers. However, the CG approaches have long been limited by the difficulty of constructing both accurate and transferable models efficiently, considering the large diversity of chemical structures of materials. Among CG force fields, Martini is the most widely used, as it retains essential chemical features while offering substantial computational efficiency. Its most recent version, Martini 3, expands chemical resolution through a much broader bead set, particularly for small molecules. However, this flexibility also complicates the mapping of organic molecules because of context-dependent rules and the lack of standardized procedures. To address this issue, we present an automated framework that builds Martini 3 models directly from SMILES (Simplified Molecular Input Line Entry System) strings by combining a curated bead dictionary with a hierarchical, rule-based algorithm and molecule-specific bonded parameters.
Our framework, Martini Mapper
\url{https://github.com/eliobaby/Martini_mapper}, generated Martini 3 models for 6,280 molecules across six chemically diverse datasets, including 1,689 systems with bond/angle parameters and additional large systems mapped at the topological level. A curated subset of 1,075 mapped structures was benchmarked using transfer free energies in hydrated octanol, hexadecane, and chloroform from water against reference data wherever available. We further examined the benchmark with structural validation via SASA, yielding good agreement with experimental and atomistic reference data. 
The workflow can also map large molecules containing up to 172 heavy atoms, 
exceeding the capabilities of existing automated approaches.
Our framework, therefore, enables systematic and scalable Martini 3 structures for high-throughput simulations relevant to drug discovery and materials design.
\end{abstract}

\section{Introduction}

\noindent

Molecular dynamics (MD) simulations were introduced in the early 1960s \cite{Rahman1964} to translate classical mechanics into predictive models of atomic motion, initially targeting simple liquids and crystalline solids. \cite{Frenkel2002}
Over the past five decades, MD has evolved into a central technique in chemical and biological physics \cite{Leach2002, Karplus2002}, enabling the resolution of molecular processes inaccessible to analytical theory or direct observation. 
By integrating empirical or quantum-derived force fields with Newtonian dynamics \cite{Unke2021}, MD provides access to atomically detailed trajectories over nanosecond to microsecond timescales \cite{Doniach1999, Freddolino2008}. 
This capability makes MD indispensable for understanding structure–function relationships in biomolecules, molecular recognition, and soft matter behavior \cite{Dror2012, Hollingsworth2012}. 
Its predictive relevance has steadily increased, particularly in contexts where experimental resolution is limited or transient intermediates dominate, such as protein conformational switching, allosteric modulation, or lipid membrane remodeling \cite{Shaw2010, Shaw2011}. 
In recent applications, MD has been leveraged to elucidate drug-binding kinetics \cite{Shaw2013}, refine cryo-EM structural models \cite{Posani2025}, and identify cryptic pockets in viral proteins \cite{Meller2023}, including the SARS-CoV-2 spike \cite{Zuzic2022, Casalino2020}.
Critically, with the increase in computational power and advancing integration techniques, MD is no longer merely descriptive; it increasingly guides experimental design, suggesting hypotheses, validating interpretations, and accelerating discovery across molecular sciences \cite{Zhou2022, DeVivo2016, Rapaport2004}.

The focus of molecular simulations has always been operating across a hierarchy of spatial and temporal resolutions, with each level balancing computational cost and chemical detail. At the all-atom (AA) level, each atom, including hydrogens, is treated as an explicit interaction site, enabling rigorous representation of directional interactions and conformational energetics. These force fields are parameterized via hybrid protocols incorporating quantum mechanical data and thermophysical observables, and are routinely benchmarked against experimental structures and time-correlation functions \cite{MacKerell1998, Cornell1995, Feig2019}. United-atom (UA) models introduce a systematic reduction by integrating nonpolar hydrogens into their parent heavy atoms, reducing the number of degrees of freedom while retaining molecular topology and core thermodynamic adherence \cite{Jorgensen1988, Gunsteren2004}.

While UA models reduce computational cost modestly, many biologically or technologically relevant processes span length- and time-scales that remain inaccessible even at this intermediate resolution. To overcome these barriers, Coarse-grained (CG) representations extend atomistic resolution reduction by mapping groups of atoms into single interaction sites or beads, thereby lowering the number of degrees of freedom and smoothing the underlying energy landscape \cite{Marrink2004, Noid2013, Voth2013}. This abstraction enables simulations of large molecular assemblies and slow collective processes, granting access to mesoscopic length and time scales beyond the reach of all-atom approaches \cite{Klein2008, Kmiecik2016, an2018development, an2020machine, bejagam2018machine}. Although the reduction in chemical specificity is an inherent trade-off, modern CG models, particularly those developed within the Martini framework have proven capable of capturing mesoscale organization, emergent dynamics, and key thermodynamic observables with remarkable fidelity. Recent advances demonstrate predictive accuracy in contexts ranging from protein–ligand binding and lipid self-assembly to biomolecular condensates \cite{Souza2020, Kjolbye2022}. When rigorously parameterized and validated against atomistic simulations or experimental benchmarks, CG models thus function not merely as simplified surrogates but as powerful, resolution-adaptive tools within multiscale simulation workflows that bridge molecular detail with emergent material and biological phenomena.

Among CG force fields, Martini is the most widely adopted due to its balance between transferability, accuracy, and computational efficiency \cite{polyply, PyLipID, Pedersen2025}. 
The original Martini 2 force field became a standard for simulating lipids, proteins, and small molecules, offering a chemically intuitive four-to-one mapping scheme \cite{Marrink2007, Monticelli2008, Vazquez2020}. However, its limited bead vocabulary often led to oversimplifications, particularly for polar and aromatic systems \cite{Alessandri2019}. Martini 3 addresses these issues by introducing a refined and expanded bead set, enhanced size resolution (tiny, small, regular), and improved mapping rules that capture chemical diversity with greater fidelity \cite{Souza2021, Grunewald2022}. 
At the same time, Martini 3 exhibits increased sensitivity to bonded interaction parametrization and molecular volume consistency, making accurate bond, angle, and structural definitions critical for stable and transferable models.
This increased resolution improves accuracy but also introduces complexity in mapping, especially for small molecules with varied functional groups \cite{Alessandri2022}. As a result, manual mapping becomes a bottleneck, motivating the development of automated, rule-based approaches to fully leverage Martini 3’s capabilities at scale \cite{Bereau2015, Potter2021, Kroon_2025, Szczuka2025}.

A range of methodologies has been developed to automate or optimize atomistic-to-coarse-grained mapping, using rule-based, graph-theoretic, and machine learning (ML) approaches.
For example, the Deep Supervised Graph Partitioning Model (DSGPM) \cite{Li2020} and MolCluster \cite{Zhong2025} use graph neural networks in supervised and unsupervised settings, respectively, to automate coarse-grained mapping as data-driven alternatives to manual approaches.
Webb \emph{et. al.}\cite{Webb2019} proposed the Graph-Based Coarse-Graining (GBCG) method, systematically generating mappings via edge contractions on molecular graphs. Zhong \emph{et. al.}\cite{Zhong2025} extended this by integrating a neural architecture with optimizer flexibility in AMOFMS, which enables both bottom-up and top-down parametrization. Potter \emph{et. al.}\cite{Potter2021} developed an automated mapping algorithm specific to the Martini force field, combining graph analysis with heuristics for ring handling and membrane partitioning benchmarks. Bereau and Kremer~\cite{Kremer2015} proposed a protocol for automatic Martini parametrization, validated on hydration and partition free energies.
Recent advances in bonded parameter automation include PyCGTOOL \cite{pycgtool}, which derives equilibrium bond and angle values directly from atomistic trajectories, and Bartender \cite{bartender}, which employs quantum mechanics–based molecular dynamics to extract Martini 3 bonded terms with improved numerical stability. These developments build upon advances in extended tight-binding (xTB) quantum chemistry methods \cite{xTB}, which enable fast and broadly applicable semiempirical atomistic simulations across diverse chemical systems. In parallel, force-field level extensions such as GFN-FF further demonstrate the feasibility of fully automated construction of molecular force-field terms across large chemical spaces \cite{genff}. Together, these approaches underscore the importance of systematic, molecule-specific bonded parametrization strategies in multiscale modeling workflows.
Wang and G{`}omez-Bombarelli~\cite{Wang2019} employed variational autoencoders to simultaneously learn CG variables and their backmapping, while Zhang \emph{et. al.}\cite{Weinan2018} introduced DeePCG, a deep learning model preserving many-body correlations. Rudzinski and Noid\cite{Rudzinski2014} provided a theoretical basis for evaluating CG mappings using iterative g-YBG theory, and Mahajan and Tang~\cite{Tang2023} presented an automated framework for polyethylenimine mapping under Martini with validation against experimental observables.

Other efforts contribute tools for CG system construction and visualization~\cite{Machado2016}, highlight statistical inconsistencies in Martini models~\cite{Voth2021}, or discuss broader applications in macromolecular modeling~\cite{Voth2018}. Although these methods advance the CG mapping landscape, key challenges persist. Most models focus on either fixed-resolution mapping or specific chemical classes, and few generalize across chemically diverse small molecules. Crucially, the expanded chemical vocabulary in Martini 3 amplifies mapping ambiguity, especially in aromatic, branched, or heteroatom-rich systems. Existing ML frameworks often depend on curated training data or fail to yield directly simulation-ready topologies \cite{Husic2020}. Moreover, while some approaches address mapping prediction, they do not integrate rule-based validation or ensure reproducibility across runs \cite{Bolhuis2024, Niki2022}. These limitations collectively motivate us to build a unified, rule-driven framework capable of mapping arbitrary molecules into Martini 3 representations with full automation, extensibility, and physical consistency.
Recent progress using the Martini 3 force field for small molecules includes Auto-MartiniM3 \cite{Szczuka2025}, which demonstrates strong scalability and performance on small chemical systems. However, systematic integration of molecule-specific bonded parameter extraction and validation across diverse chemical spaces remains an open challenge.
%Although Auto-MartiniM3 exhibits excellent scalability and efficiency, its applicability remains restricted to molecules containing fewer than approximately 20-25 heavy atoms, beyond which the automated mapping and parametrization procedures lose reliability.
Furthermore, it is to note that although the Martini 3 model construction is supported by comprehensive guidelines and well-documented best practices \cite{Souza_book, martini3_tutorials}, applying these conventions systematically across large and chemically diverse datasets remains non-trivial in high-throughput settings.

In this work, we present an automated framework that generates Martini 3 coarse-grained models of small molecules with a range of heavy atoms from 2 to 172,
directly from canonical Simplified Molecular Input Line Entry System (SMILES) strings.
The motivation of Martini\_Mapper is to formalize and automate the application of these existing rules into a reproducible and extensible workflow suitable for large-scale molecular screening.
The framework combines a curated bead-mapping dictionary with a hierarchical, rule-based mapping algorithm and integrates molecule-specific bonded parameters derived from xTB-based ensemble sampling. The algorithm automatically identifies rings, side chains, and functional groups through structural analysis, followed by prioritized hierarchical bead assignment that enforces molecular symmetry.
Across six chemically diverse datasets, Martini Mapper generates 6,280 mapped topologies, including 1,689 systems for which bonded parameters were derived from xTB-based molecular dynamics sampling.
To evaluate predictive fidelity, we benchmark the generated models against experimentally measured transfer free energies in multiple solvent systems (hydrated octanol, hexadecane, and chloroform), together with structural validation via solvent-accessible surface area (SASA) comparisons against atomistic reference structures. These validations demonstrate reasonable agreement across chemically diverse systems.
%A hierarchical mapping strategy then assigns beads by prioritizing ring systems, followed by non-ring systems and chain fragments, enforcing molecular symmetry.
Our framework produces GROMACS-compatible coordinate and topology files in a fully reproducible and simulation-ready form, thereby enabling scalable high-throughput model construction. 
%The algorithm and the complete set of more than 6000 mapped itp/top files are available at the Martini Mapper \url {https://github.com/eliobaby/Martini_mapper}. To evaluate its predictive fidelity, we benchmarked the generated models against experimental partition (water–octanol) free energy, which demonstrates excellent performance for a wide variety of target structures within Martini parametrizations. 
Finally, we report current limitations and outline future directions of Martini\_Mapper. %toward extending chemical space and incorporating adaptive parametrization strategies. 
In essence, our model establishes a reproducible foundation for data-driven coarse-grained modeling, facilitating applications in drug discovery, polymer–drug assembly, and biomolecular condensates.

\section{Algorithmic Framework for Automated Coarse-Grained Mapping}

\begin{figure}[h!]
%\centering
\includegraphics[width=0.8\textwidth]{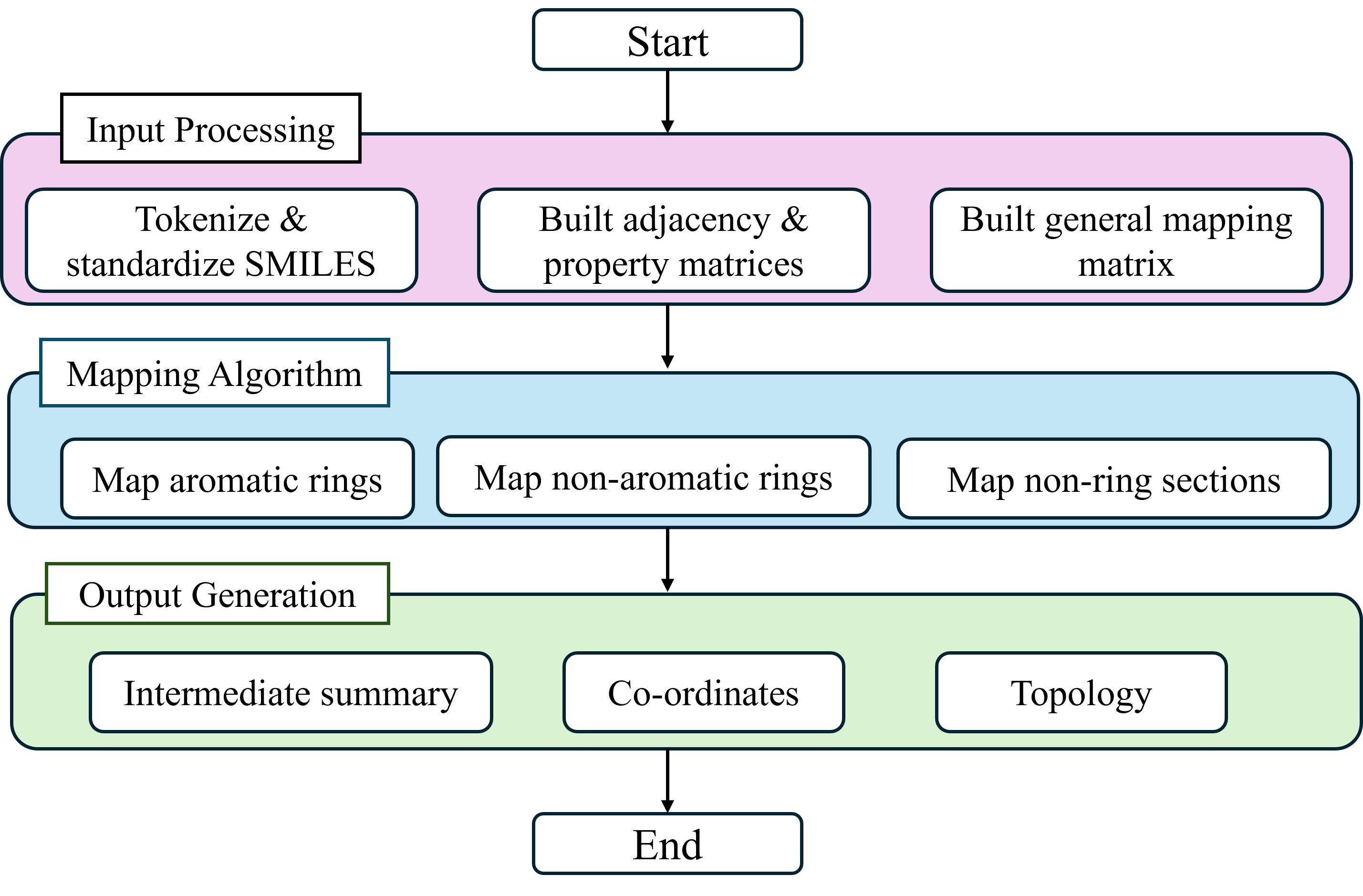}
\caption{The flowchart of our automated coarse-grained mapping pipeline. The process begins with SMILES input, proceeds through preprocessing (tokenization, graph construction, and mapping array generation), applies the hierarchical bead assignment algorithm, and outputs simulation-ready coordinate (\texttt{.gro}) and topology (\texttt{.itp}) files.}
\label{pipeline}
\end{figure}

The automated algorithmic framework is a fragment-based method developed for mapping atomic fragments to coarse-grained beads. The first step in this framework is constructing a bead dictionary, which serves as a reference that links specific molecular fragments to predefined coarse-grained bead types. 
Each bead entry corresponds to a chemically meaningful unit, such as an alkyl chain, an aromatic ring fragment, or a functional group. The mapping process consists of three key steps: input processing, mapping, and output generation, which is shown in the flowchart in Figure~\ref{pipeline}.
The input processing involves pre-processing the SMILES string of a molecule.\cite{weininger1988smiles} 
This string encodes the full atom-level structure of the molecule, including its connectivity, ring closures, and branching patterns. 
From this representation, the framework extracts the topological information necessary to identify chemically distinct fragments and prepare required matrices (\emph{e.g.} adjacency/property matrices) for mapping.
Once the input is processed, the framework applies a hierarchical rule-based mapping algorithm to assign beads to different parts of the molecule. 
The mapping rules are organized so that the mapping starts with the most structurally constrained regions, which in our framework typically ring systems, and then proceeds to non-ring fragments, including chains and side groups.
This layered approach allows the framework to ultimately generate Martini~3-compatible coarse-grained models (including the summary, coordinates, and topology) in a fully deterministic and reproducible manner.
We will explain each of these steps with specific examples in detail below.

\subsection{Building Literature-based Building Block Table (LBBT)}

To construct the initial bead dictionary, we integrated two complementary sources from the Martini 3 framework and compiled them into a literature-based building block table (LBBT).
The first source is the list of all the 90 small molecules
reported in the Martini small molecule dataset \cite{Alessandri2022}, where each entry corresponds to a chemically defined substructure derived from validated coarse-grained models.
%Examples include: (a) CH$_{3}$CH$_{2}$COOH mapped to bead P2, (b) CH$_{3}$COOH mapped to SP2, (c) a –CH$_{2}$OH group bonded to a trifluoromethyl unit mapped to TP1d, \red{(d) aromatic N=N mapped to TN1a, (e) methyl-pyrrole ring fragments mapped to bead TN1, and (f) non-ring alkyl fragments C(C)(C)C mapped to bead SC1.} 
These fragment-level entries are chemically precise with respect to the influence of neighboring atoms and topological features. 
The second source, that we use is the supplementary Table 24 from the Martini 3 force field \cite{Souza2021}, which lists default bead assignments for generic chemical groups without contextual detail. 
Here, the examples include: (a) alcohol mapped to P1, (b) carboxylic acid to P2, (c) phenol to N6, (d) linear alkane to C1, etc. 
While this table lacks fragment specificity, it covers the most common functional groups.

%Moreover, we have expanded the LBBT by including more new fragment-bead mapping, as detailed in Section 2.4.

%\red{As a third source, we incorporated the Martini 3 small-molecule database, which provides a curated set of 90 small molecules with human-defined bead mappings and accompanying topologies that have been tested within the Martini 3 framework \cite{Alessandri2022}.
%In contrast to generic functional-group defaults, these molecule-level models encode chemically significant context (e.g., ring membership, conjugation, aromaticity, and substitution patterns) directly in the validated mapping choices.
%Examples include: (a) TN1a assignments for aromatic N=N fragments, (b) methyl-pyrrole ring fragments mapped to bead TN1, and (c) non-ring alkyl fragments C(C)(C)C mapped to bead SC1, which are handled as distinct topological environments relative to linear alkanes.}
% We must add a reference for Grunewald2025 CGSMILES
As a third source, we incorporated the benchmark dataset reported in the Gr{\"u}newald study \cite{Grunewald_2025}, which introduces a line-notation designed to represent coarse-grained models across resolutions.
%and, importantly for our purposes, analyzes a benchmark set of 407 molecules from the Martini force field together with their coarse-grained descriptions.
This dataset is chemically significant because it systematically aggregates many validated Martini mappings in machine-readable form, spanning diverse functional groups and bead types, thereby broadening coverage beyond fragment lists and generic group defaults.
From this dataset, we added fragment-to-bead correspondences that recur across mapped molecules, including: (a) sulfonamide-like fragments mapped to P4, which we represent with multiple chemically distinct sulfonamide fragments, (b) cyanamide mapped to P4d, (c) carbon disulfide mapped to C6, and (d) carbon dioxide mapped to SC6.
These three sources were then merged to create a unified dictionary, LBBT, containing a total of 254 fragments. %by including all distinct entries.
%All these additions improve LBBT coverage to fragments that are underrepresented in purely functional-group default tables, while remaining grounded in published Martini mappings~\cite{Kroon_2025, Alessandri2022, Alessandri2019}.}

\subsection{Preprocessing of Molecular Structures from SMILES}

\begin{figure}[h!]
%\centering
\includegraphics[width=0.9\textwidth]{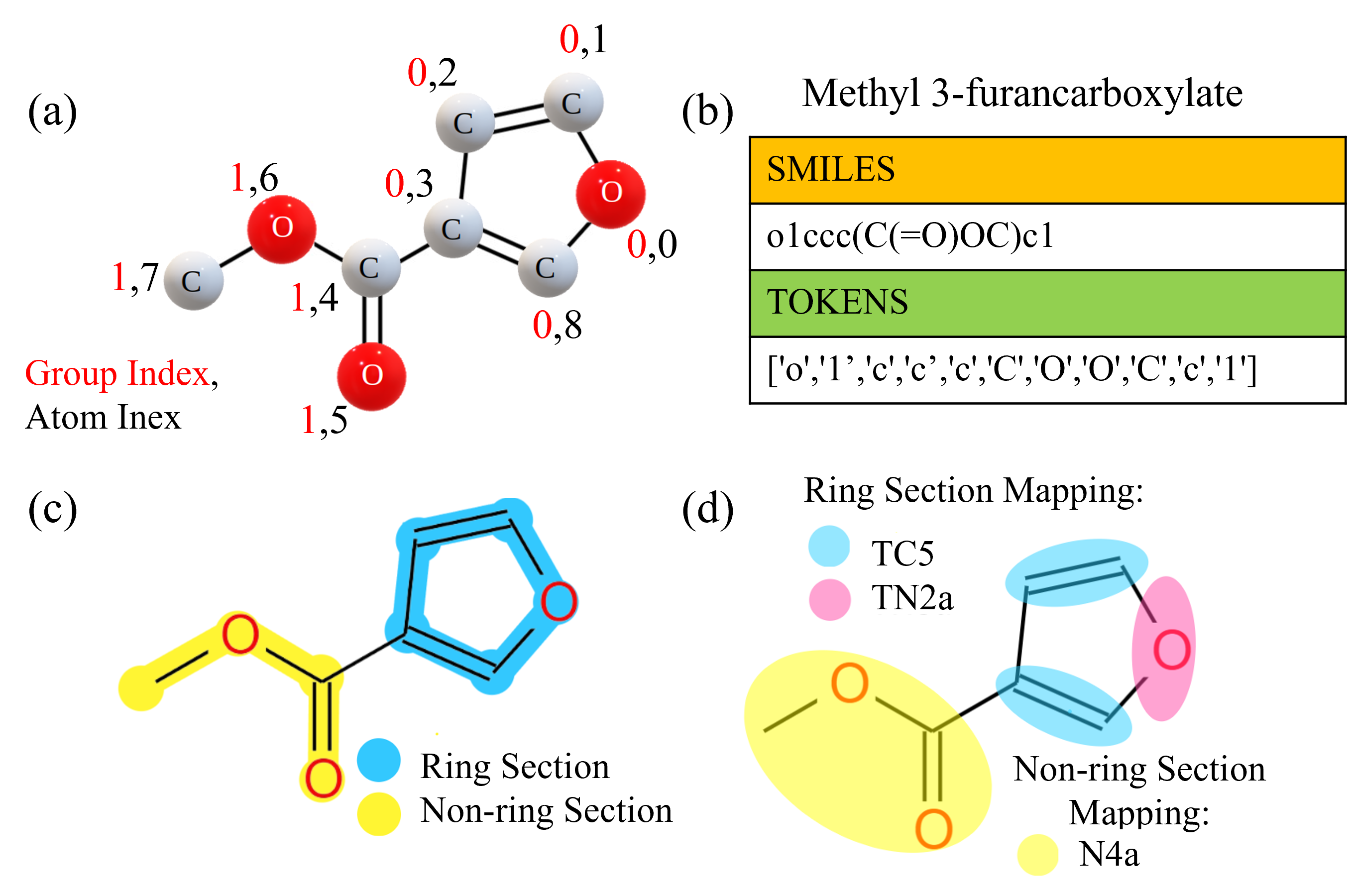}
\caption{ (a) Chemical structure of methyl 3-furancarboxylate. Note hydrogen atoms are not shown for clarity, but the number of hydrogen atoms attached to heavy atoms are counted (see Table 1).
(b) Tokenization of the canonical SMILES string (\texttt{o1ccc(C(=O)OC)c1}) into atomic and structural symbols.
Mapping Scheme: (c) Partitioning of the molecule into ring (blue) and non-ring (yellow) sections according to the algorithm. (d) Final bead assignments: the aromatic ring is mapped into TC5 and TN2a beads, while the methoxycarbonyl group is mapped into an N4a bead.}
\label{bromo_token}
\end{figure}

As described before, our automated CG mapping framework begins with a SMILE representation of the target molecule. 
This preprocessing stage is performed in the following steps that transform a one-dimensional text-based input into a structured data model suitable for rule-based bead assignment within the Martini 3 force field beads. 

\begin{enumerate}[label=(\roman*)]
    \item \textbf{SMILES tokenization and standardization:} The canonical SMILES string is parsed into a discrete sequence of tokens, where each token represents either an atom or a structural modifier such as a ring closure index. Tokenization ensures that equivalent structures yield identical token sequences, while standardization removes stereochemical and isotopic annotations not required at CG resolution. This reduction simplifies the mapping space, allowing consistent pattern recognition and minimizing ambiguity during rule matching. For illustration, the methyl 3-furancarboxylate molecule (\texttt{o1ccc(C(=O)OC)c1}) 
    is tokenized into an ordered sequence corresponding to a five-membered aromatic ring containing one oxygen atom and a methoxycarbonyl (linear) side group
     (Figure~\ref{bromo_token}).

\begin{table}[h!]
\centering
\begin{tabular}{c}
% Left table: Property Matrix
\textbf{Property Matrix} \\
\begin{tabular}{|c|c|c|c|c|c|}
\hline
\textbf{Atom} & \textbf{Element} & \textbf{Is} & \textbf{Ring} & \textbf{Is } & \textbf{Hydrogen} \\
\textbf{Index} &  & \textbf{in Ring?} & \textbf{Status} & \textbf{Edge?}  &  \textbf{Count} \\
\hline
0 &	O  & T  & 2 & F & 0\\
1 &	C  & T  & 2 & F & 1\\
2 &	C  & T  & 2 & F & 1\\
3 &	C  & T  & 2 & F & 0\\
4 &	C  & F  & 0 & T & 0\\
5 &	O  & F  & 0 & T & 0\\
6 &	O  & F  & 0 & T & 0\\
7 &	C  & F  & 0 & T & 3\\
8 &	C  & T  & 2 & F & 1\\
\hline
\end{tabular}
\\
\textbf{Connectivity Matrix} \\
% Right table: Connectivity Matrix
\begin{tabular}{|c|c|c|c|c|c|c|c|c|c|}
\hline
 & \textbf{0} & \textbf{1} & \textbf{2} & \textbf{3} & \textbf{4} & \textbf{5} & \textbf{6} & \textbf{7} & \textbf{8} \\
\hline
\textbf{0} & 0 & 1.5 &	0 &	0 &	0 &	0 &	0 &	0 &	1.5 \\
\textbf{1} & 1.5 &	0 &	1.5 & 0 & 0 & 0 & 0 & 0 & 0 \\
\textbf{2} & 0 & 1.5 &	0 &	1.5 & 0 & 0 & 0 & 0 & 0 \\
\textbf{3} & 0 &	0 &	1.5 &	0 &	1 &	0 &	0 &	0 &	1.5 \\
\textbf{4} & 0 &	0 &	0 &	1 &	0 &	2 &	1 &	0 &	0 \\
\textbf{5} &	0 &	0 &	0 &	0 &	2 &	0 &	0 &	0 &	0 \\
\textbf{6} &	0 &	0 &	0 &	0 &	1 &	0 &	0 &	1 &	0 \\
\textbf{7} &	0 &	0 &	0 &	0 &	0 &	0 &	1 &	0 &	0 \\
\textbf{8} & 1.5 &	0 &	0 &	1.5 &	0 &	0 &	0 &	0 &	0 \\
\hline
\end{tabular}
\\
\end{tabular}
\caption{Property and connectivity matrices of methyl 3-furancarboxylate. In the property matrix, the ring status, 2, 1, 0 represents aromatic rings, non-aromatic rings, and non-ring fragments, respectively. Hydrogen count values represent how many hydrogen atoms are attached to the corresponding heavy atoms. In the connectivity matrix, 0, 1.0, 1.5, and 2.0 represent no bonds, single bonds, bonds in aromatic rings, and double bonds.}
\label{tab:property_connectivity}
\end{table}

    \item \textbf{Construction of adjacency and property matrices:} From the tokenized representation, two primary data structures are built. The \emph{property matrix} encodes atom-level attributes such as element type, aromaticity, ring membership, whether the atom lies at a fragment boundary (edge status), and hydrogen counts, which is used to distinguish fragments that share the same SMILES, e.g., separating primary, secondary, and tertiary amide/amine environments, local bonding environment such as whether an oxygen is in an ether (no hydrogen attached) or a hydroxyl (OH). The \emph{connectivity matrix} encodes bond topology, with single bonds represented as 1.0, bonds in aromatic rings as 1.5, and double bonds as 2.0. For methyl 3-furancarboxylate, the five carbons are flagged as aromatic in a pentane ring, while the non-ring linear chain is identified as a terminal substituent (see Table~\ref{tab:property_connectivity}).
    \item \textbf{General mapping array generation:} The adjacency and property information are combined into a general mapping array, a hierarchical grouping of atoms into structural sections such as aromatic rings, non-aromatic rings, and non-ring fragments, encoded as ``2", ``1", and ``0" in ``Ring Status", accordingly. Each entry includes atom indices, element types, connectivity context, and whether the atom is at the edge of a fragment.
    For methyl 3-furancarboxylate, the four carbons and one oxygen are flagged as aromatic in a five-membered ring, while the non-ring linear chain is identified as a terminal substituent (see Table~\ref{tab:property_connectivity}).
    The resulting mapping array partitions the molecule into two chemically distinct fragments: the five-membered aromatic ring (atoms 0–3, 8) and the non-ring substituent (atoms 4–7). After defining these fragment boundaries,  the framework specifies how atoms across fragments are connected, such as the single bond between atom 3 of the ring and atom 4 of the substituent, and atoms within a section are connected using inner connections, such as the connections between the atoms inside the ring, preserving complete molecular connectivity.
    For outer connections (connecting to other groups/sections), the bond entry is broken down into three parts: the index of the group in which the foreign atom resides (Outer Group), the index of the atom within that section (Outer Atom), and the bond order between the atoms (Outer Bond). For inner connections, only the bonded atom’s index within the section (Inner Atom) and the bond order (Inner Bond) are needed. This structure cleanly encodes both local and cross-section connections, as shown in Table~\ref{mapping_array}.

\begin{table}[h!]
\centering
\scriptsize
\setlength{\tabcolsep}{3pt}
\renewcommand{\arraystretch}{1.1}
\begin{tabular}{|c|c|c|c|c|c|c|c|c|c|c|}
\hline
\textbf{Group} & \textbf{Atom} & \textbf{Atom Index}  & \textbf{Outer} & \textbf{Outer} & \textbf{Outer} & \textbf{Inner} & \textbf{Inner} & \textbf{Inner} & \textbf{Inner}   \\
\textbf{Index} & \textbf{Index} & \textbf{within group} & \textbf{Group 1} & \textbf{Atom 1} & \textbf{Bonds 1} & \textbf{Atom 1} & \textbf{Bond 1} & \textbf{Atom 2} & \textbf{Bond 2}  \\
\hline
\multicolumn{10}{|c|}{\textbf{Ring Group (Group 0)}} \\
\hline
0 & 0 & 0  & -- & -- & -- & 1 & 1.5 & 4 & 1.5  \\
0 & 1 & 1  & -- & -- & -- & 0 & 1.5 & 2 & 1.5  \\
0 & 2 & 2  & -- & -- & -- & 1 & 1.5 & 3 & 1.5  \\
0 & 3 & 3  & 1 & 0 & 1 & 2 & 1.5 & 4 & 1.5  \\
0 & 8 & 4  & -- & -- & -- & 0 & 1.5 & 3 & 1.5  \\
\hline
\multicolumn{10}{|c|}{\textbf{Non-Ring Group (Group 1)}} \\
\hline
1 & 4 & 0  & 0 & 3 & 1 & 1 & 2 & 2 & 1  \\
1 & 5 & 1  & -- & -- & -- & 0 & 2 & -- & --  \\
1 & 6 & 2  & -- & -- & -- & 0 & 1 & 3 & 1  \\
1 & 7 & 3  & -- & -- & -- & 2 & 1 & -- & --  \\
\hline
\end{tabular}
\caption{Mapping array showing internal and external connections and edge status for methyl 3-furancarboxylate.}
\label{mapping_array}
\end{table}

\end{enumerate}

In our framework, the tokenization process is entirely automated, enabling consistent treatment of molecules of varying architectures from polycyclic scaffolds to functionalized aromatic systems.

\subsection{Hierarchical Mapping Strategy}

Following the initial processing of the input SMILES string, the core of our framework is a hierarchical mapping strategy that translates the atomistic structure into a coarse-grained model. This procedure uses the main mapping matrix. 
It sorts atoms into distinct ring and non-ring sections. A key principle of our approach is that bead assignment is performed dynamically.
As the algorithm analyzes each section, groups of atoms that match a rule in our dictionary are immediately assigned a bead type, and this information is used to guide the mapping of subsequent, connected fragments.
The algorithm is broadly divided into two main stages: the mapping of ring and non-ring structures. The separation of ring (blue) and non-ring (yellow) sections is illustrated in Figure~\ref{bromo_token}(c).
The reason behind the preference to map rings is explained below. 

\begin{itemize}
    \item \textbf{Establishing a Rigid Foundation:} Ring systems, particularly aromatic ones, are the most structurally rigid parts of a molecule. By mapping these stable foundations first, we establish a set of fixed anchor points. The more flexible non-ring sections can then be mapped in the context of these rings.
    \item \textbf{Preventing Complications with Lone Atoms:} Many non-ring sections consist of single atoms (e.g., a hydroxyl oxygen) attached to a ring. Attempting to map these lone atoms first would be problematic, as their correct bead assignment almost always depends on merging them with the larger ring structure they are attached to. By mapping the ring sections first, these lone atoms are naturally merged and assigned to a defined structure.
\end{itemize}

\begin{figure}[htb!]
    \centering
    \includegraphics[width=0.9\linewidth]{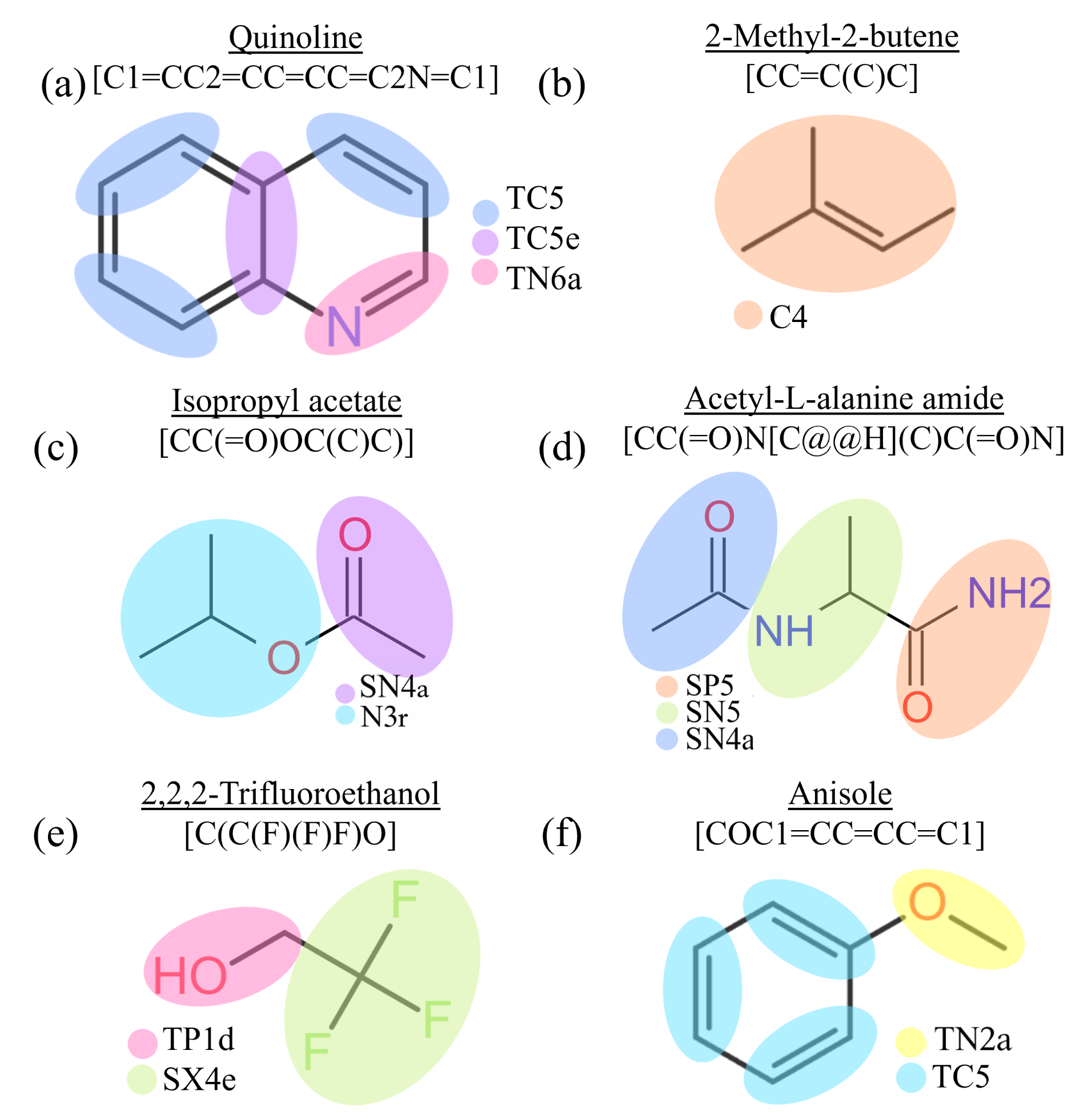}
    \caption{Mapping of representative molecules. (a) Quinoline is treated by the algorithm as two ring sections with a ring-fusion point, which is mapped to TC5e, while the remaining parts of the molecule are mapped to TC5 and TN6a according to their chemical structure (C=C to TC5 and C=N to TN6a)
    (b) 2-Methyl-2-butene, is mapped to C4, here path length, $l$ is equal to $3$.
    (c) Isopropyl acetate is treated by the algorithm as a single non-ring fragment, and is mapped into two beads (SN2 and N2), when $l > 3$.
    (d) Acetyl-L-alanine amide is mapped into three beads (SP5, SN5, and SN4a), for $l > 3$.
    (e) 2,2,2-Trifluoroethanol is mapped into two beads (TP1d, SX4e).
    (f) Anisole is mapped into TN2a and TC5.
    The coloring of the beads is only tailored to each image.}
    \label{non_ring_fragment}
\end{figure}

\subsubsection{Mapping Ring Structures}
As mentioned before, the mapping of ring systems in coarse-grained models requires a systematic approach to preserve their structural and chemical characteristics. Because aromatic and non-aromatic rings exhibit different bonding patterns and levels of rigidity, the algorithm applies a priority-based procedure tailored to each case. We illustrate the mapping of methyl 3-furancarboxylate in Figure~\ref{bromo_token}(c) and (d).
The steps below outline this hierarchy, ensuring that the most critical features are captured first before completing the mapping of the full ring.

\begin{enumerate}
    \item \textbf{Ring Fusion Points:} The algorithm first identifies and maps any atoms that are part of more than one ring system. These fusion points are the most constrained atoms in the molecule, and mapping them first provides a stable scaffold for the rest of the section. They are typically grouped with a neighbor and assigned a specialized fused-ring bead type (e.g., TC5, TC5e). We provided an example of this fusion ring in Figure~\ref{non_ring_fragment}(a) by mapping the Quinoline molecule (SMILES: \texttt{C1=CC2=CC=CC=C2N=C1}), where the purple bead TC5e is the fusion point.

    \item \textbf{Double Bonds in Non-aromatic Ring:} Next, atoms involved in double bonds are prioritized. In aromatic rings,  all atom connections share a delocalized $\pi$-bonding equivalent to alternating single/double bonds. However, double bonds can still exist in non-aromatic rings. Double bonds are more rigid than single bonds, making them the next most important structural feature to anchor. They are typically mapped as two-atom beads. If any of the atoms in the double bond are connected to a lone external section, that section is often grouped with the double-bonded atoms to form a larger, three-atom bead that captures the entire functional group. 

    \item \textbf{Ring atoms with a single-atom non-ring neighbor:} The algorithm then proceeds to map ring atoms that are connected to single-atom non-ring sections. The ring atom, its unmapped ring neighbor, and the external lone atom are grouped into a three-atom bead. This step ensures small functional groups (like hydroxyls or halogens) are treated as a single chemical unit.

    \item \textbf{Remaining Unmapped Atoms:} Finally, any unmapped atoms that do not belong to the previously mapped categories (ring fusion points, double bonds, or ring atoms with external substituents) are grouped to complete the ring structure. For aromatic rings, these are typically paired into two-atom beads (e.g., TC5). The resulting bead assignments for the example are shown in Figure~\ref{bromo_token}(d) (TC5 and TN2a for the ring; N4a for the methoxycarbonyl side chain). For non-aromatic rings, they are often grouped into three-atom beads (e.g., SC3), a common representation in coarse-graining that effectively captures the geometry of CH$_2$–CH$_2$–CH$_2$ fragments. 
\end{enumerate}

\subsubsection{Mapping Non-Ring Structures}
After all ring systems have been mapped, every remaining unmapped atom belongs to a non-ring section of size at least 2 heavy atoms. 
The mapping of these remaining larger non-ring sections is guided by the Martini 3 path length (\textit{l}) constraint, which is defined as the maximum number of consecutive covalent bonds spanned within a single coarse-grained bead and must not exceed three (i.e., $ l \leq 3$).
At this stage, all lone-atom sections have already been accounted for by being merged into the rings they were attached to.

\begin{itemize}
    \item \textbf{Single-Bead Mappable Fragments (\textit{l} $\le$ 3):} The algorithm first tests each non-ring section to determine its length. If the longest path between any two atoms in the fragment is three bonds or fewer, it is mapped to a single bead. A canonical signature is generated based on its structure to find the correct bead type in the dictionary.
    \begin{itemize}
        \item For \textbf{linear chains}, the signature is a simple string of atom and bond types (e.g., `CC=CC').
        \item For \textbf{branched fragments}, the signature is defined by the central atom followed by the signatures of each branch (e.g., `C(C)(C)(CC)'). This ensures a unique representation regardless of atom ordering.
    \end{itemize}
    
\end{itemize}

\begin{itemize}
    \item \textbf{Complex Fragments Requiring Splitting ($ l > 3$):} Now, if a fragment is too large, it must be broken down using a recursive splitting strategy. Here, recursive indicates that the splitting procedure is applied iteratively: the fragment is divided once, the resulting pieces are evaluated, and any sections that still violate the path length constraint are further subdivided until all fragments comply.
    \begin{itemize}
        \item For long \textbf{linear chains}, the algorithm prioritizes efficiency by partitioning the chain into the largest possible mappable units, which are typically four-atom fragments ($l = 3$). For example, an octane molecule would be split into two four-carbon beads.
        \item For complex \textbf{branched structures}, a more adaptive strategy is used. The algorithm identifies the two `edge' atoms (terminals) with the shortest path between them. This path is broken off to form the first new section, and the remaining atoms form the second. This is repeated until all resulting fragments are small enough to be mapped by a single bead. In the edge case where multiple edge atoms are equidistant (e.g., in a cross-like structure like neopentane), they are grouped and broken off together. A representative example of Isopropyl acetate and Acetyl-L-alanine amide (see Figure \ref{non_ring_fragment} (c),(d),(e)), is shown here, which requires splitting into multiple beads. Example (c) specifically shows that when that non-ring section is broken apart, it forms 2 parts that are both 1-bead mappable, and one is linear and the other one is branched. Example (d) illustrates a case where a single split does not fully resolve the section into single-bead mappable fragments. After the initial division, one of the resulting fragments still exceeds the path-length limit, so the algorithm recursively applies a second split. Example (e) further illustrates the edge case where multiple edge atoms are equidistant. In this example, the three fluorine edge atoms are broken off together with their central carbon atom.
    \end{itemize}

   \item \textbf{Resolving Functional-Group Ambiguities Using Hydrogen Counts:} 
    In some cases, the SMILES pattern of a fragment alone is not sufficient to uniquely identify the functional group, because the same heavy-atom connectivity can correspond to multiple chemistries depending on substitution. This occurs for motifs such as amines, imines, amides, and for oxygenated groups that are ambiguous without protonation state (e.g., carboxylic acid vs.\ ester, acetal/ketal vs.\ hemiacetal/hemiketal, or diol-like motifs). 
    To resolve these ambiguities without explicitly querying neighboring beads, the algorithm uses the \emph{hydrogen counts} stored in the property matrix (see Table~\ref{lbbt_table}). Hydrogen counts allow us to infer how many substituent (\emph{R}) groups are attached to a heteroatom within the fragment, which determines its substitution class and thus the correct bead assignment.
    \begin{itemize}
        \item \textbf{Amide/amine substitution (primary/secondary/tertiary):} 
        for an N-centered fragment, the number of attached hydrogens distinguishes whether the nitrogen is primary (NH$_2$, two H), secondary (NH, one H), or tertiary (no H). For example, the heavy-atom pattern \texttt{NC(=O)} corresponds to an amide, but the N--H count differentiates a primary amide (\texttt{H$_2$NC(=O)R}), a secondary amide (\texttt{HNC(=O)R$_2$}), and a tertiary amide (\texttt{NC(=O)R$_3$}). In our dictionary, these are mapped to distinct Martini bead types: primary amides are assigned to \textbf{P5}, secondary amides to \textbf{P3}, and tertiary amides to \textbf{P3a}. This distinction is chemically meaningful because it changes polarity and hydrogen-bonding capability, and therefore can change the appropriate Martini bead type.
        \item \textbf{Carboxylic acid vs. ester:} 
        the fragment pattern \texttt{C(=O)O} is shared by both carboxylic acids and esters. If the oxygen in this motif carries one hydrogen (O--H), the group is a carboxylic acid (\texttt{C(=O)OH}) and is assigned to \textbf{P2}; if the oxygen has zero hydrogens, it must be substituted (\texttt{C(=O)OR}) and is therefore an ester, assigned to \textbf{N4a}. This resolves the ambiguity using only atom-local information.
        \item \textbf{Acetal/ketal vs. hemiacetal/hemiketal vs. diol:} 
        acetal-like motifs contain a carbon bonded to two oxygens (e.g., \texttt{C(OR)(OR)}). The hydrogen counts on the two oxygens indicate whether either oxygen is a hydroxyl: if one oxygen has an O--H (one H) while the other is substituted (zero H), the motif is a hemiacetal/hemiketal (\texttt{C(OH)(OR)}) and is assigned to \textbf{P2}; if both oxygens have zero hydrogens, the motif is a full acetal/ketal (\texttt{C(OR)(OR)}) and is assigned to \textbf{N4a}. In the limiting case where both oxygens have O--H (each one H), the motif corresponds to a hydrate/diol-like environment (\texttt{C(OH)(OH)}), which is chemically distinct from acetals and is assigned to \textbf{P4}.
        \item \textbf{Ether vs.\ alcohol:}
        oxygen-centered fragments can also be ambiguous from heavy-atom connectivity alone, because an \texttt{O} atom may represent either an ether oxygen (\texttt{R--O--R}) or a hydroxyl oxygen (\texttt{R--O--H}). The hydrogen count resolves this directly: an ether oxygen has zero attached hydrogens, implying two R groups, whereas an alcohol oxygen has one attached hydrogen, implying one R group. In our dictionary, ether-like environments can be assigned to \textbf{N2a} or \textbf{N3a/r} depending on the surrounding chemical context (e.g., aromatic vs.\ aliphatic substitution and local polarity), while alcohol-like environments can be assigned to \textbf{N6} or \textbf{P1} depending on context (e.g., phenolic vs.\ aliphatic alcohol).
    \end{itemize}
    
\end{itemize}

%%% BLue means Grunewald; green means 2D/Kaggle/Bereau; Red means Original 90/BBT

\begin{table}[h!]
\centering
\caption{Mapping of Functional Classes and Groups to LBBT Beads. The table shows bead type reassignment; bead size resolution (R, S, T) follows standard Martini 3 definitions.
The building blocks of Alcohol, Ether, Carboxylic acid, and Ester are mapped to Martini 3 beads, identical to those from Martini 3 datasets
~\cite{Souza2021,Alessandri2022}. The amine and imine groups are mapped to the beads, same as reported by the Gr{\"u}newald dataset~\cite{Grunewald_2025}.
}

%\begin{tabular}{|p{2cm}|p{6cm}|p{1.8cm}|>{\centering\arraybackslash}m{6cm}|}
\begin{tabular}{|l|l|c|c|}
\hline
\textbf{Functional Class} & \textbf{Functional Group} & \textbf{LBBT Bead} & \textbf{Representative Molecules}\\
\hline
Amine & Primary amine & N6d & 
\includegraphics[width=3cm]{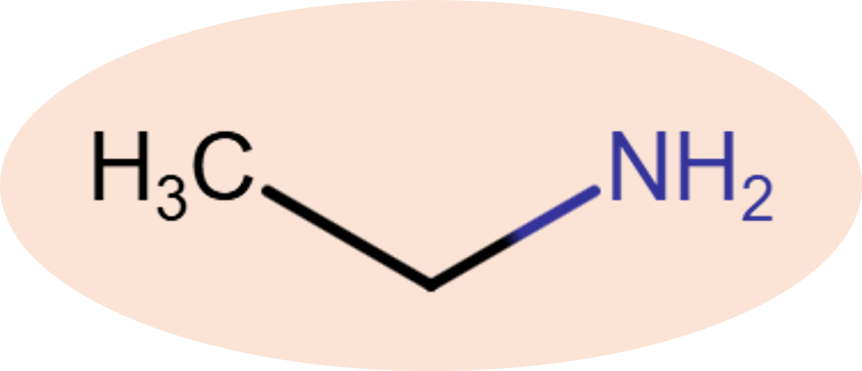} \\
 &  (e.g., CH$_3$--CH$_2$--NH$_2$) & & Ethylamine\\% \cite{Grunewald_2025}} \\
\hline
Amine & Secondary amine 
& N5 & 
\includegraphics[width=4cm]{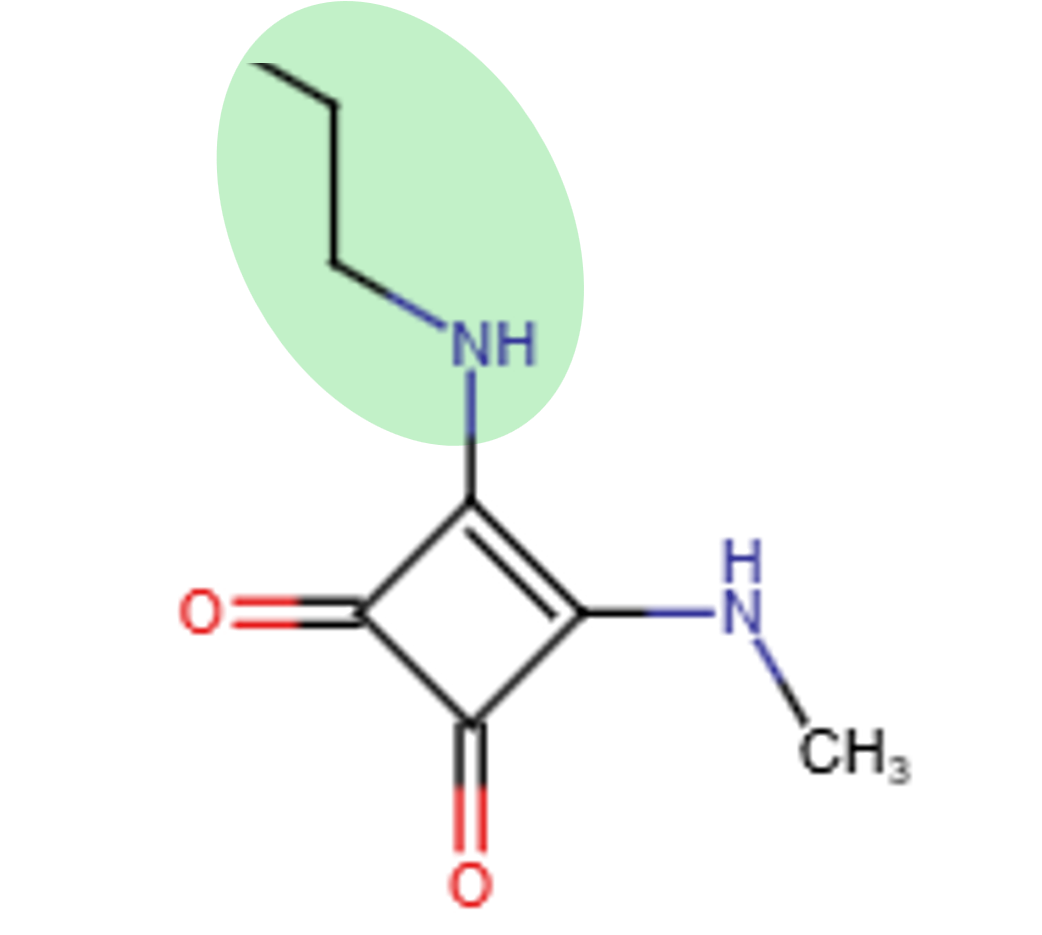}  \\
 & (e.g., --NH--CH$_2$--)  & & Squaramide\\% \cite{Grunewald_2025}} \\
\hline
Amine & Tertiary amine & N3a & 
\includegraphics[width=2.5cm]{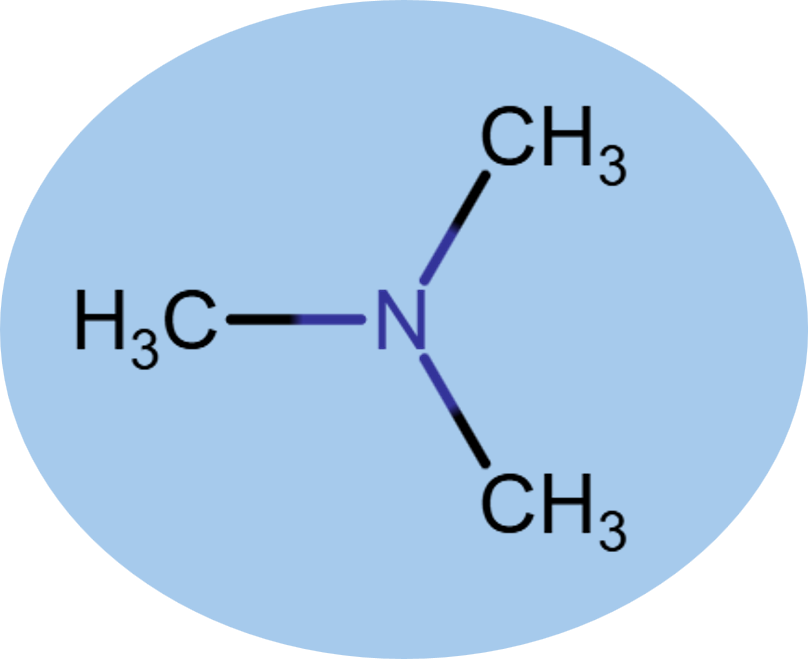} \\
 &  (e.g., N--(CH$_3$)$_3$)  & & Trimethylamine\\% \cite{Grunewald_2025}} \\
\hline
Amide & Primary amide & P5 & 
\includegraphics[width=3cm]{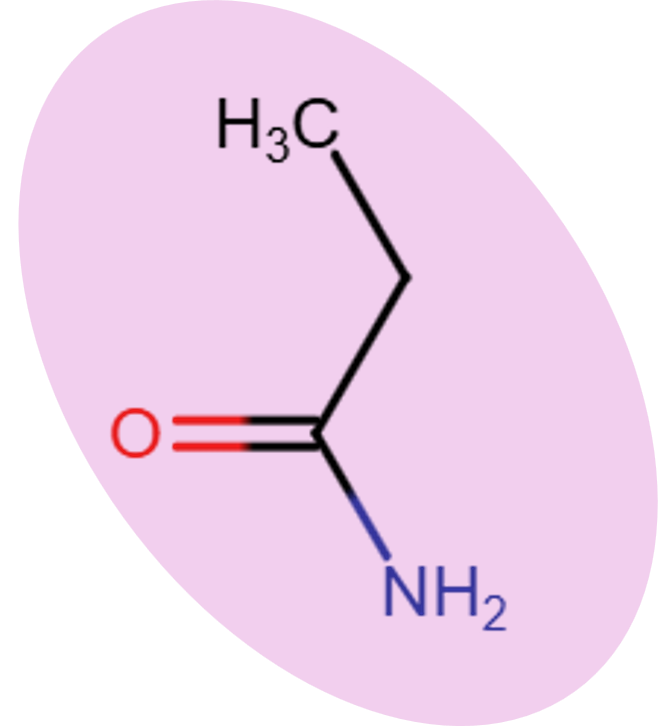} \\
 & (e.g., --CONH$_2$)   & & Propanamide\\% \cite{Grunewald_2025}} \\
 \hline
 Amide & Secondary amide & P3 & 
\includegraphics[width=3cm]{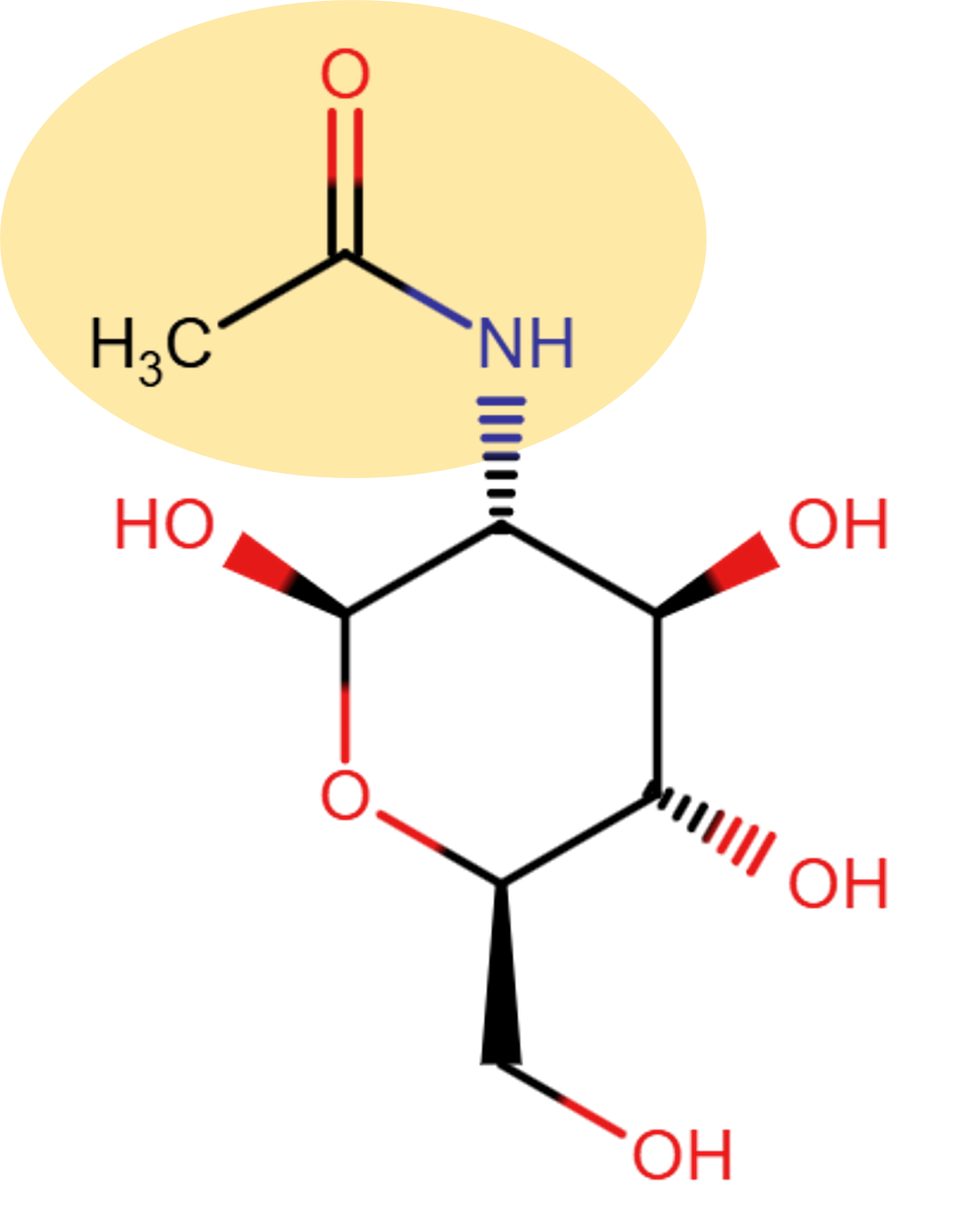} \\
 & (e.g., --CONH--CH$_3$)   & & N--acetylglucosamine\\% \cite{Grunewald_2025}} \\
\hline
\end{tabular}
\label{lbbt_table}
\end{table}

\begin{table}[h!]
\ContinuedFloat
\centering
\caption[]{(continued)}
%\begin{tabular}{|p{2cm}|p{6cm}|p{1.8cm}|>{\centering\arraybackslash}m{6cm}|}
\begin{tabular}{|l|l|c|c|}
\hline
\textbf{Functional Class} & \textbf{Functional Group} & \textbf{LBBT Bead} & \textbf{Representative Molecules}\\
 \hline
Amide & Tertiary amide & P3a &  
\includegraphics[width=3cm]{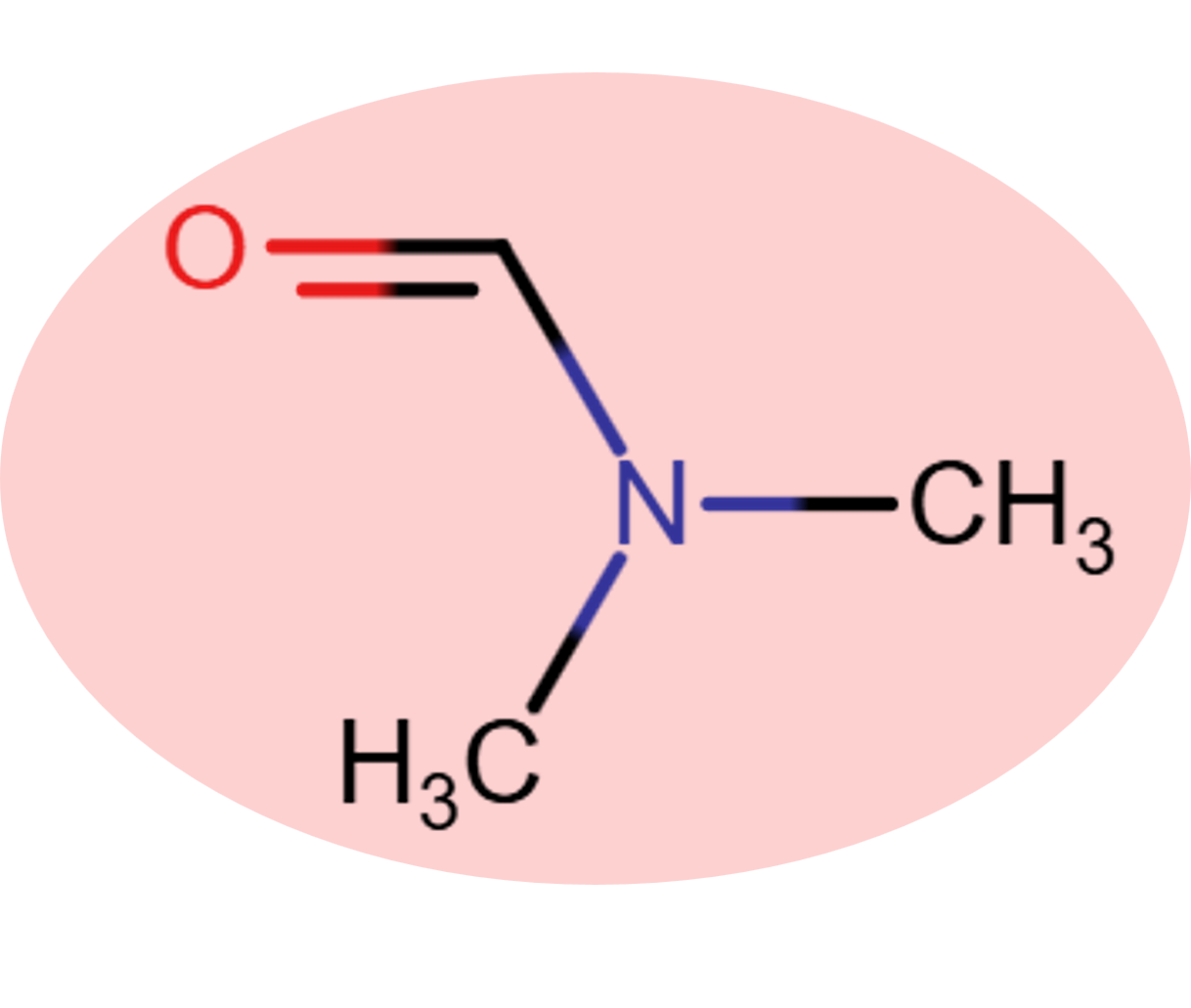} \\
 &   (e.g., O=C--N(CH$_3$)$_2$) &  & N,N-dimethylformamide \\% \cite{Grunewald_2025}} \\
 \hline
 Imine & Primary imine & N2  & \includegraphics[width=4cm]{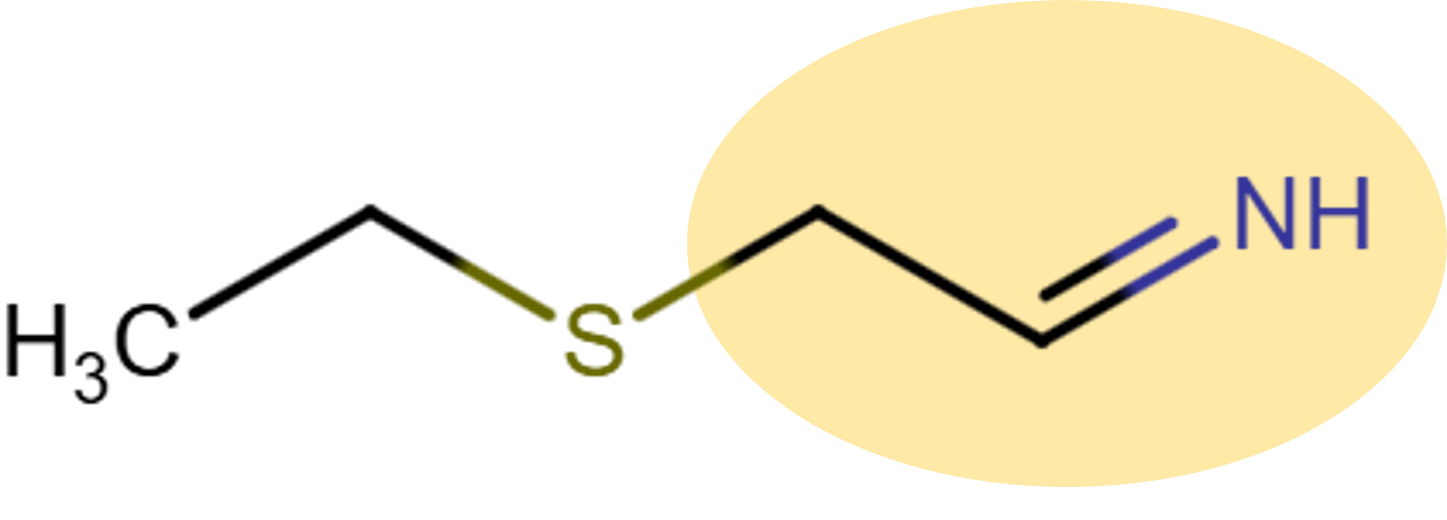} \\
  &   (e.g., --CH=NH)  & & (ethylthio)methanimine \\ %\cite{2Da, 2Db, Kaggle, Bereau2015}} \\
\hline
%N-(3-chloropropyl)methanimine
Imine & Secondary imine  & N1a &  \includegraphics[width=4cm]{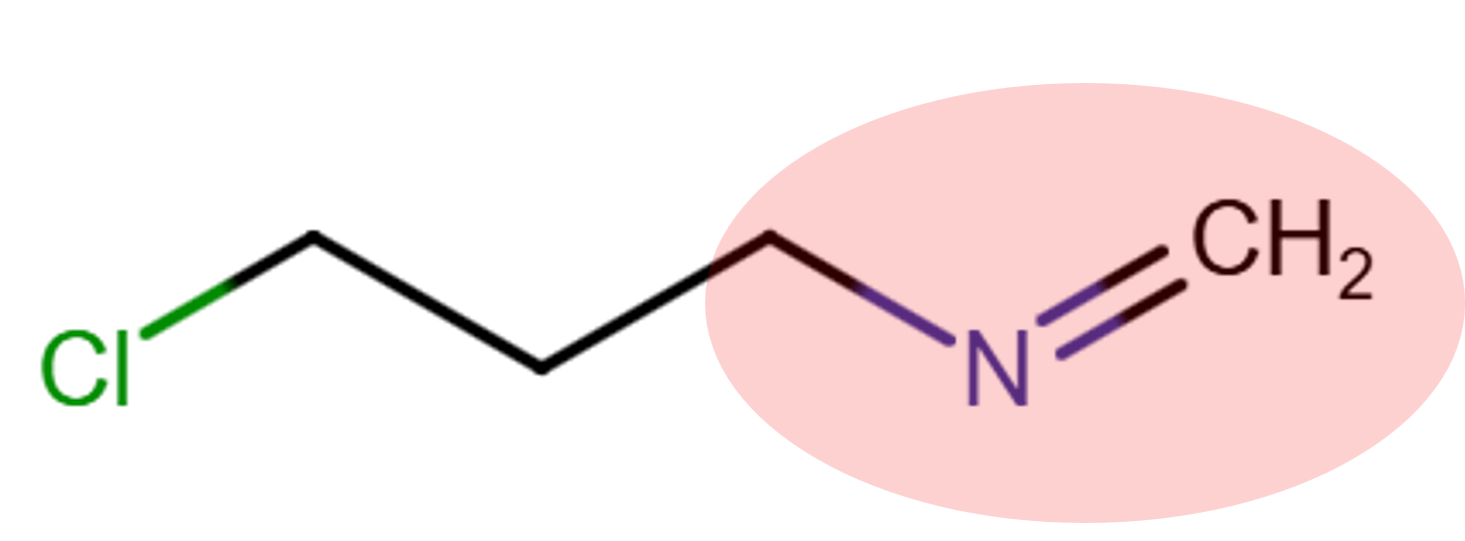}\\ 
& (e.g., --CH$_2$--N=CH$_2$) & & N-(3-chloropropyl)methanimine \\% \cite{2Da, 2Db, Kaggle, Bereau2015}} \\
\hline
Acetal-like & Acetal / Ketal & N4a & \includegraphics[width=2.5cm]{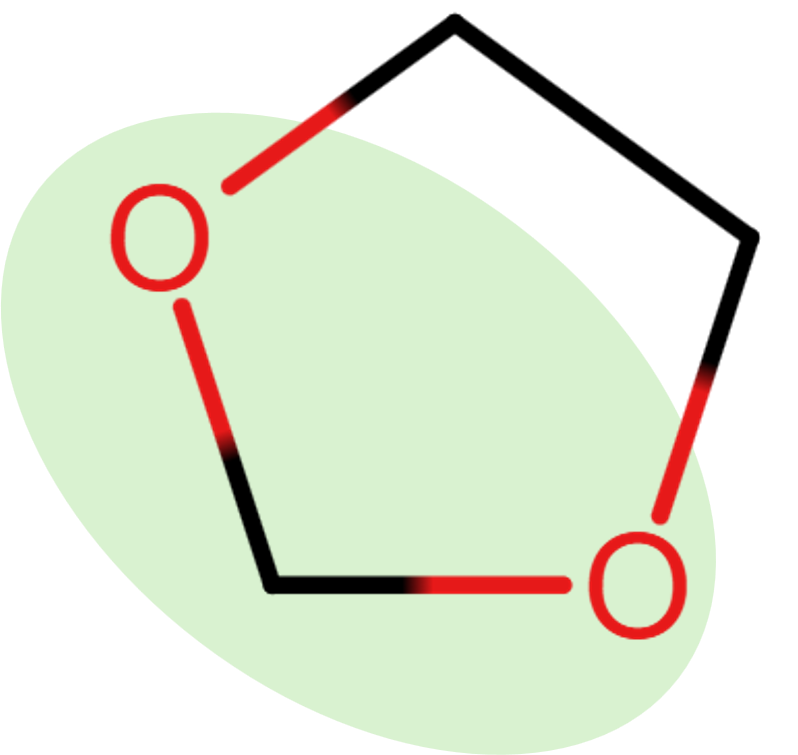} \\
& (e.g., CH$_2$(O--)$_2$)  & & 1,3-dioxolane \\%\cite{2Da, 2Db, Kaggle, Bereau2015}} \\
 \hline
 Acetal-like & Hemiacetal / Hemiketal & P2  & \includegraphics[width=4cm]{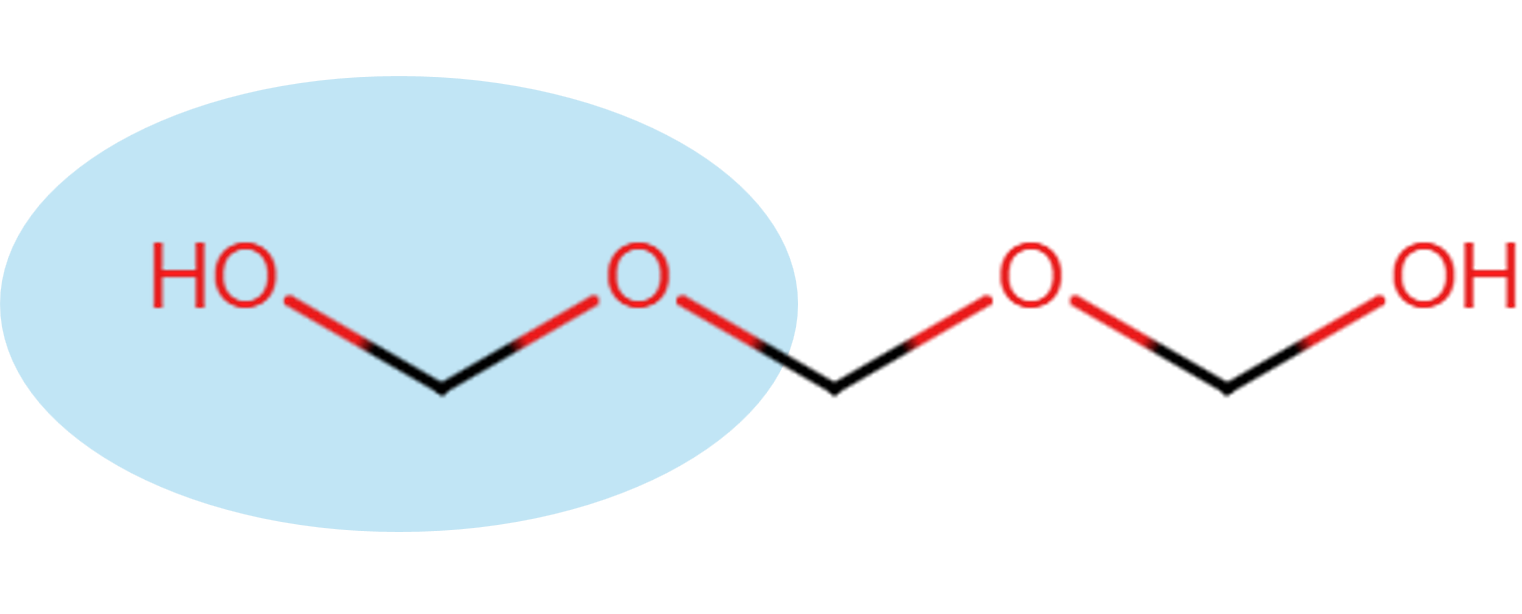}  \\
& (e.g., OH-CH$_2$--O--)  & & (Methylenebis(oxy))dimethanol \\% \cite{Grunewald_2025}} \\
\hline
Acetal-like & Diol  & P4  & \includegraphics[width=4cm]{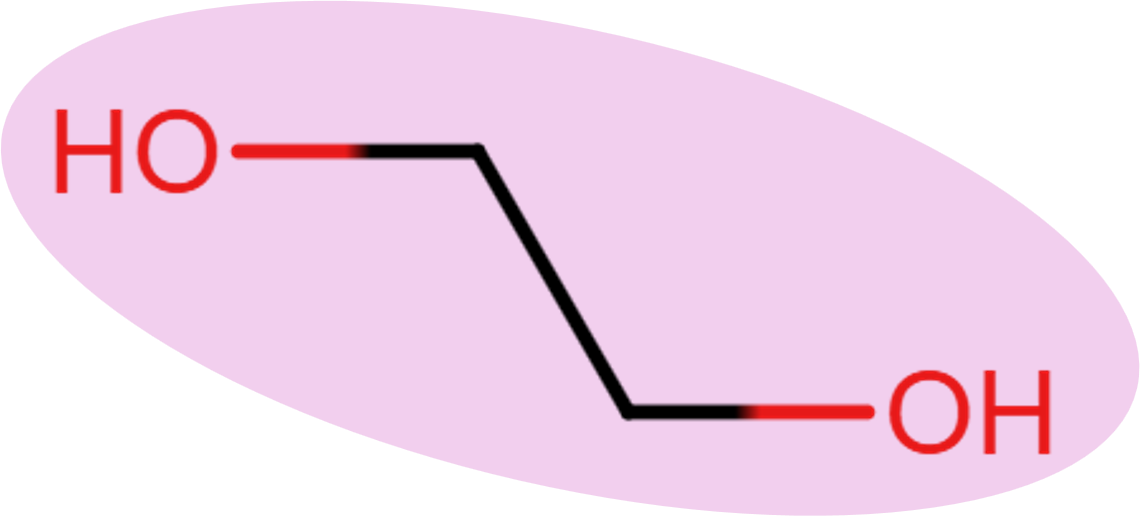} \\
& (e.g., OH--CH$_2$--CH$_2$--OH)  & & ehtylene-glycol \\% \cite{Grunewald_2025}} \\
\hline
 Alcohol / Ether & Alcohol & P1 & \includegraphics[width=3.5cm]{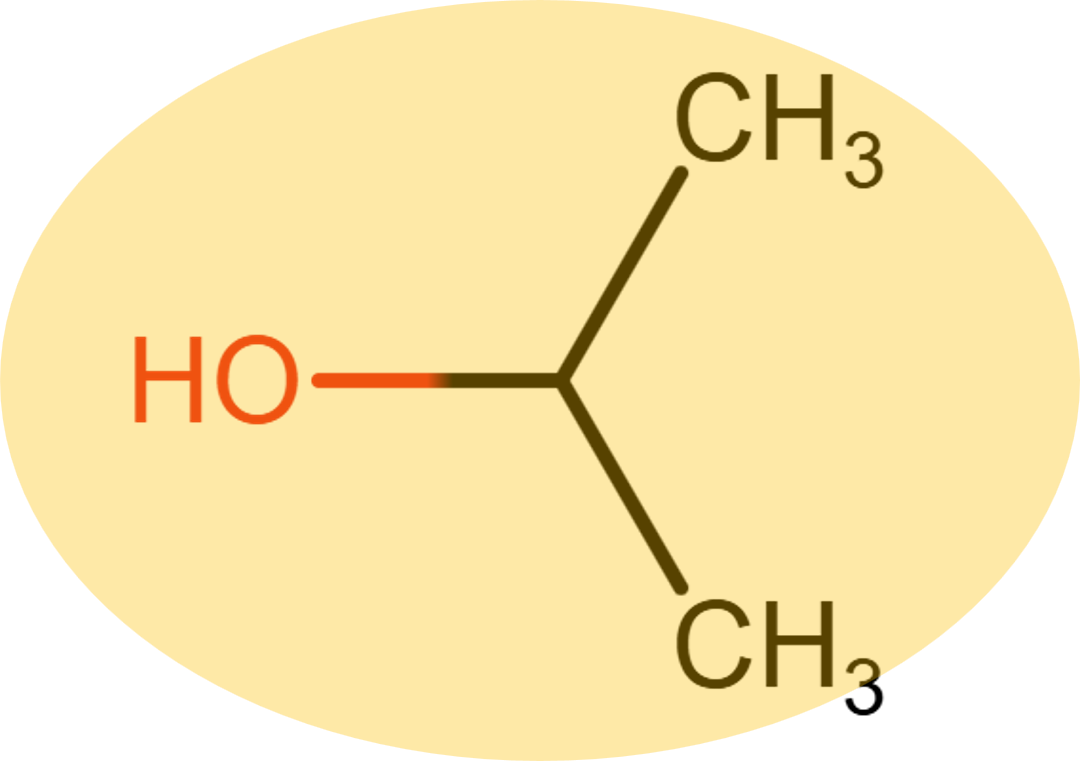}  \\
 &  (e.g., OH--CH(CH$_3$)$_2$)  & & Isopropanol \\% \cite{Souza2021, Alessandri2022}} \\
\hline
\end{tabular}
\end{table}

\begin{table}[h!]
\ContinuedFloat
\centering
\caption[]{(continued)}
%\begin{tabular}{|p{2cm}|p{6cm}|p{1.8cm}|>{\centering\arraybackslash}m{6cm}|}
\begin{tabular}{|l|l|c|c|}
\hline
\textbf{Functional Class} & \textbf{Functional Group} & \textbf{LBBT Bead} & \textbf{\red{Representative Molecules}}\\
\hline
Alcohol / Ether & Ether & N3r & \includegraphics[width=4cm]{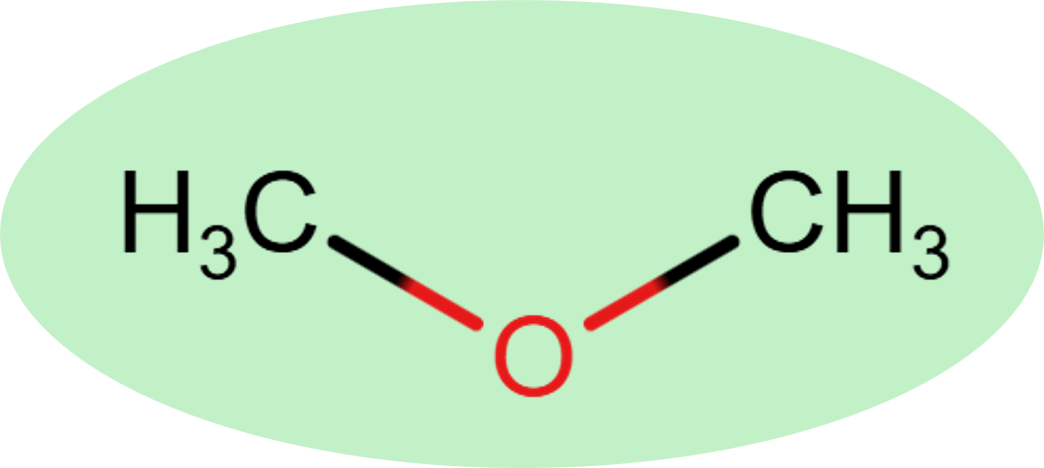} \\
&  (e.g., CH$_3$--O--CH$_3$)   & & Dimethyl ether\\% \cite{Souza2021, Alessandri2022}} \\
\hline
Carboxyl & Carboxylic acid & P2 & \includegraphics[width=3cm]{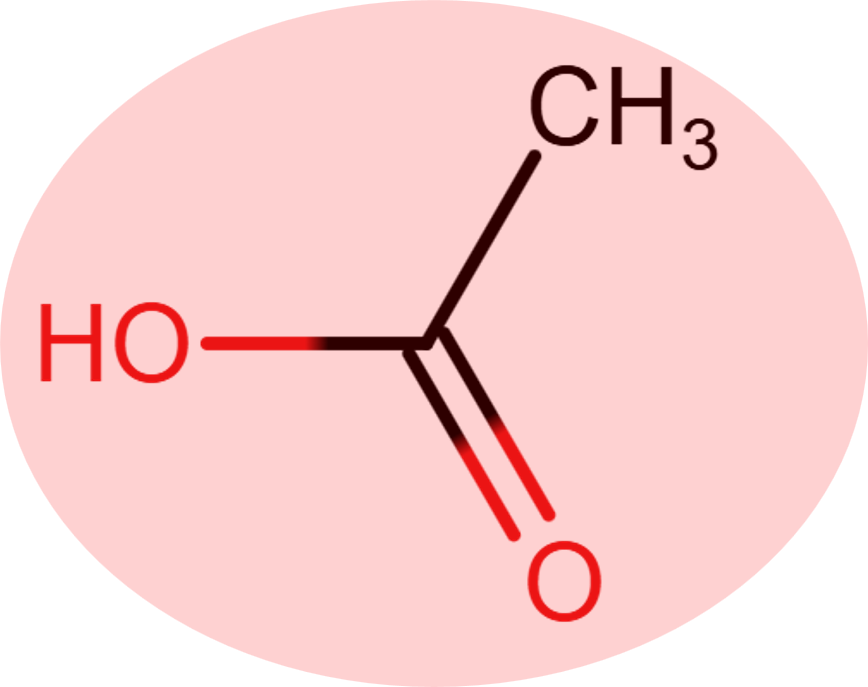} \\
& (e.g., CH$_3$--COOH)   & & Acetic Acid \\% \cite{Souza2021, Alessandri2022}} \\
\hline
Carboxyl & Ester & N4a & \includegraphics[width=3cm]{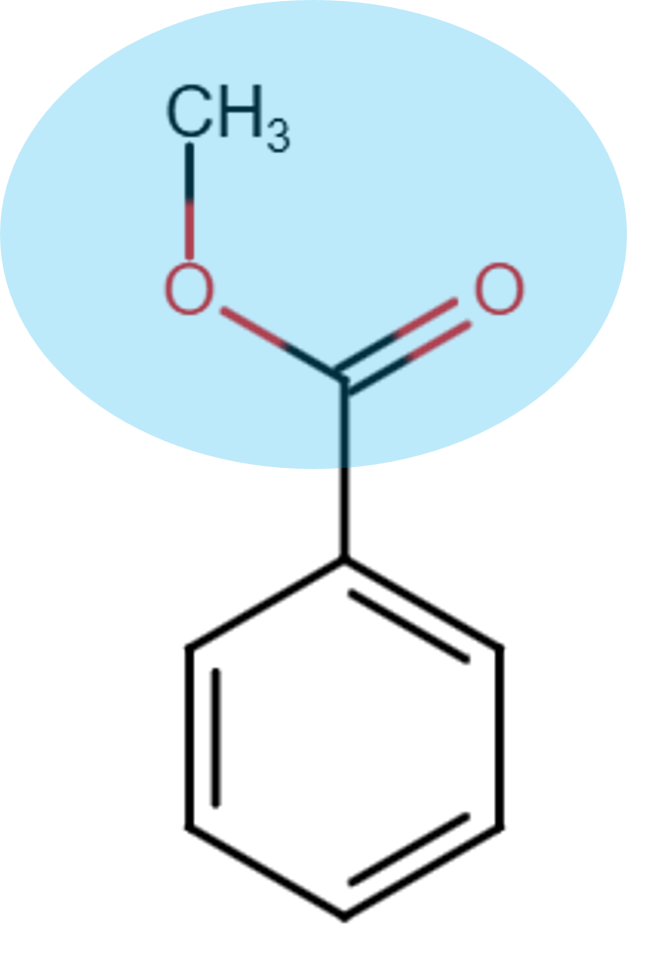} \\
&  (e.g., --COO--CH$_3$)  & & Methyl Benzoate \\% \cite{Souza2021, Alessandri2022}} \\
\hline
\end{tabular}
\end{table}

\subsection{Output Generation}
After this final step, every atom in the original molecule has been assigned to a coarse-grained bead, and the mapping is ready for output generation. It is essential to develop an algorithm that not only performs bead mapping but also generates bead coordinates and bond lengths, thereby producing a consistent coordinate set and bond topology.
To begin this step, the initial three-dimensional coordinates of the all-atom structure are obtained using RDKit \cite{RDKit}, which generates a conformer from the input SMILES string.
Once a molecule is generated from its SMILES string and mapped to beads, each bead corresponds to a specific set of atoms.
The center of geometry of these atoms is then computed to define the bead’s coordinate. This ensures bead positions correspond to the all-atom geometry. 
The bond length between two connected beads is then obtained as the Euclidean distance between their respective centers of geometry.

\subsection{Bond and Angle Potential Parameters from xTB-Based Ensemble Sampling}

Bonded parameters are derived from constant temperature sampling of an all-atom reference trajectory generated using an extended tight-binding workflow \cite{xTB}, followed by the approach taken in Bartender \cite{bartender}. The objective is to obtain equilibrium bond lengths and angles, as well as their corresponding force constants, from statistically meaningful fluctuations rather than from a single conformer.
Starting from a SMILES representation, a three-dimensional structure is generated using RDKit \cite{Bento2020} with ETKDG \cite{ETKDG} embedding, followed by the Universal Force Field optimization \cite{Tosco2014}. This structure is subsequently optimized using the GFN-FF Hamiltonian within the xTB framework. GFN-FF is selected because it provides computational efficiency and stable sampling suitable for high-throughput mapping workflows, while maintaining chemically consistent geometries.
Following optimization, a finite-temperature classical molecular dynamics simulation is performed using xTB at $T = 300$ K in the NVT ensemble. 
This produces 1000 frames sampling conformational fluctuations around the optimized structure. The simulation parameters are:

\begin{itemize}
    \item Temperature: 300 K
    \item Total simulation time: 10 ps
    \item Integration timestep: 0.1 fs
    \item Trajectory output interval: 10 fs
\end{itemize}

%\paragraph{\red{Mapping to Coarse-Grained Coordinates}}

Each trajectory frame is mapped to a CG representation using the COG of the atoms assigned to each bead, including all explicitly added hydrogens. 
The use of COG ensures consistency with the Martini mapping criterion adopted in Martini 3 and avoids mass-weighting biases that would otherwise alter fluctuation statistics.
The details of the bond and angle equilibrium values and their corresponding force constant are detailed in Section S1 of the Supporting Information (SI).

\subsection{Ability to Map Molecules with $>$ 170 Heavy Atoms}

To evaluate the capability of the developed framework, we examined the size distribution of molecules that could be automatically mapped. Specifically, we assessed how well the algorithm performs for molecules containing different numbers of heavy atoms. Heavy atoms refer to all non-hydrogen atoms (e.g., C, N, O, S, halogens). For this analysis, we used \cite{Bereau2015, Grunewald_2025, Kaggle, 2Da, 2Db, TPCN2024} representative datasets of increasing chemical diversity and size: the Bereau dataset \cite{Bereau2015}, the Gr{\"u}newald dataset \cite{Grunewald_2025}, the Kaggle \(\log P\) dataset \cite{Kaggle}, and the 2D \(\log P\) Molecular Benchmark \cite{2Da, 2Db}. To further probe the scalability of the algorithm, we also analyzed the TPCN (Terpenoids Content Network) database.\cite{TPCN2024} The TPCN is a curated collection of over six thousand naturally occurring terpenoids derived from 1,254 plant species across 156 families.

\begin{figure}
    \centering
\includegraphics[width=0.65\linewidth]{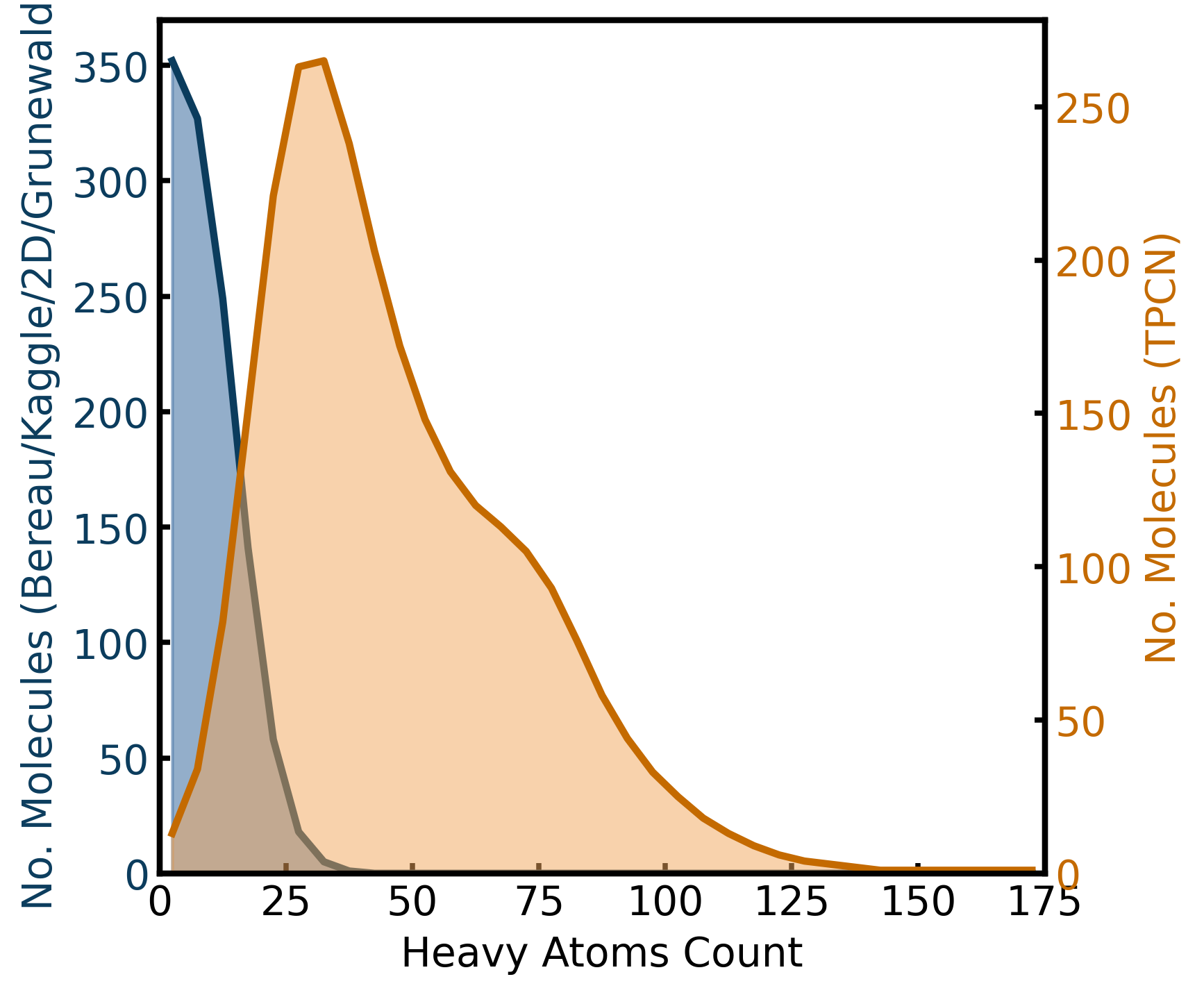}
    \caption{ The distribution of successfully mapped molecules with varying heavy-atom counts is shown in blue for all three datasets: Bereau \cite{Bereau2015}, Kaggle \cite{Kaggle}, 2D \cite{2Da}, and Gr{\"u}newald \cite{Grunewald_2025} dataset, and in orange separately for the TPCN dataset \cite{TPCN2024}.}
    \label{bbt_example}
\end{figure}

As shown in Figure~\ref{bbt_example}, our algorithm can successfully map not only the smaller organic molecules from the Bereau, Kaggle, 2D, and Gr{\"u}newald datasets, typically containing up to 16 heavy atoms, but also the larger, more intricate molecules from the TPCN dataset, which span a broad range of sizes and include species exceeding 170 heavy atoms. This demonstrates the robustness and transferability of the framework for large-scale molecular applications. The accuracy of the coarse-grained models of these molecules has been evaluated in Section 4.

\section{Simulation Details}
\label{simu}

To assess the reliability and physical consistency of the generated coarse-grained models, we performed two levels of validation: (i) thermodynamic integration (TI) calculations for transfer free energies, and (ii) equilibrium stability simulations in the NPT ensemble (see results in Section \ref{success}.
In this work, we used the TI \cite{Kirkwood} to compute the free energy from water, hexadecane, chloroform, and octanol using Martini 3 force field parameters. For each successfully mapped molecule, the pipeline-generated coordinate (\texttt{.gro}) and topology (\texttt{.itp}) files were directly incorporated into GROMACS for TI calculations in four separate simulation environments: pure water, hexadecane, chloroform, and hydrated 1-octanol with 0.2 M of water. Within the Martini 3 framework, these species are modeled using the W bead, a four-bead C1 chain, an SC2–SC2–SP1 representation, and the X2 bead type, as detailed in ref \cite{Souza2021}. To reproduce experimentally relevant conditions, the octanol phase was modeled in its water-saturated form by incorporating water at a mole fraction of 0.2 \cite{Vanikka}.
Simulations employed the stochastic dynamics (\texttt{sd}) integrator with a timestep of 20 fs (\texttt{dt} = 0.020 ps). Each $\lambda$-window was simulated for 4 ns, with 20 intermediate $\lambda$ values for van der Waals decoupling (\(0.0 \rightarrow 1.0\)) and simultaneous electrostatic decoupling, defined as:

$\lambda$ = 0,0.05,0.1,0.15,0.2,0.25,0.3,0.35,0.4,0.45,
              0.5,0.6,0.65,0.7,0.75,0.8,0.85,0.9,0.95,1

Soft-core potentials were applied with parameters \(\texttt{sc-power} = 1\), \(\texttt{sc-alpha} = 0.5\), and \(\texttt{sc-r-power} = 6\) to ensure smooth decoupling and avoid singularities.
All simulations used the Verlet cutoff scheme with \(r_{\mathrm{vdw}} = r_{\mathrm{coul}} = 1.1\) nm, a shifted potential for both van der Waals and Coulomb interactions (\texttt{Potential-shift-verlet}).
%and a relative dielectric constant \(\epsilon_r = 15\).
The temperature and pressure of the simulation have been set to 300 K using the v-rescale thermostat (\(\tau_t = 1.0\) ps) and at 1.0 bar using the Parrinello–Rahman barostat (\(\tau_p = 2.0\) ps, compressibility \(= 4.5 \times 10^{-5}\ \mathrm{bar}^{-1}\)).
Free energy differences $\Delta G$ between the coupled and decoupled states for both water and octanol bath were computed separately from the TI output. From their free energy differences, we computed $ log P$ as,

\begin{equation}
    log\ P = \frac{\Delta G _{OW}}{2.303RT}
\end{equation}

For water/hexadecane, and water/chloroform, we simultaneously obtained the respective difference in free energy.
To ensure the robustness of our thermodynamic integration protocol, we first benchmarked the methodology against the reference Martini 3 bead hydration and solvation free energies reported in the SI of Souza \emph{et. al.} \cite{Souza2021} (see Figure S1 of Section S2 in the SI). For each bead type, free energies of transfer in both water and octanol were reproduced within statistical uncertainty, yielding a near-perfect linear correlation with published values (R$^{2}$ = 0.99). This validation step confirms that the present TI setup faithfully reproduces the Martini 3 reference data before extending it to small-molecule systems. 

For the equilibrium stability simulations, the integration timestep was set to 20 fs and ran for 10 ns. If any molecule failed to complete 10 ns simulations, the respective one has been further simulated with 10 fs and 5 fs or lower timestep. The systems were maintained at 300 K temperature and 1.0 bar pressure with the same thermostat and barostat parameters as the TI simulation.

\section{Results and Performance Evaluation}

In this section, we discuss the performance of our framework on chemically diverse and publicly available four molecular databases (the Bereau dataset, the Kaggle \(\log P\) dataset, the 2D \(\log P\), and TPCN Molecular database) by first evaluating its ability to generate a valid Martini~3 mappings. Among them, we computed $\Delta G_{OW}$ or \(\log P\) values and benchmarked them against experimental references for the first three datasets due to affordable computational cost.
%In total, we evaluated its performance across a total of unique molecules drawn from these three chemically diverse and publicly available datasets.
These comprise a chemically diverse library of natural and synthetic compounds enriched in intricate topologies, including fused and overlapping ring systems, polycyclic scaffolds, and rings of uncommon sizes. It provide an ideal stress for a rule-based mapping procedure, ensuring that the framework can accommodate challenging chemistries encountered in real-world applications.

\noindent
The complete workflow of our framework was evaluated as a three-stage process:
\begin{enumerate}
\item generation of GROMACS-specific \texttt{.itp} and \texttt{.gro}. The \texttt{.itp} file contains bead types, bond, and angle information, with constraints with bond force constant exceeding 20,000 kJ/mol. The \texttt{.gro} file contains the bead coordinates.  
\item successful passage through GROMACS MD Engine, and
\item completion of TI simulations in water and octanol, and give $\Delta G$ difference between water and octanol as output. 
%without numerical instabilities or bead–bead overlap artifacts.
\end{enumerate}

We consider a successful mapping only when it can pass through all these three steps, and is termed ``working".
In total, our framework successfully generated coarse-grained models of 6,280 molecules (``Working”) over 8,850 molecules. 
%corresponding to an overall success rate of 77.8\%. 
The remaining 2,570 molecules (``Not Working”) failed primarily due to chemical fragments that were not yet represented in the mapping dictionary, rather than breakdowns in mapping logic or simulation stability.

\begin{figure}
    \centering
    \includegraphics[width=1\linewidth]{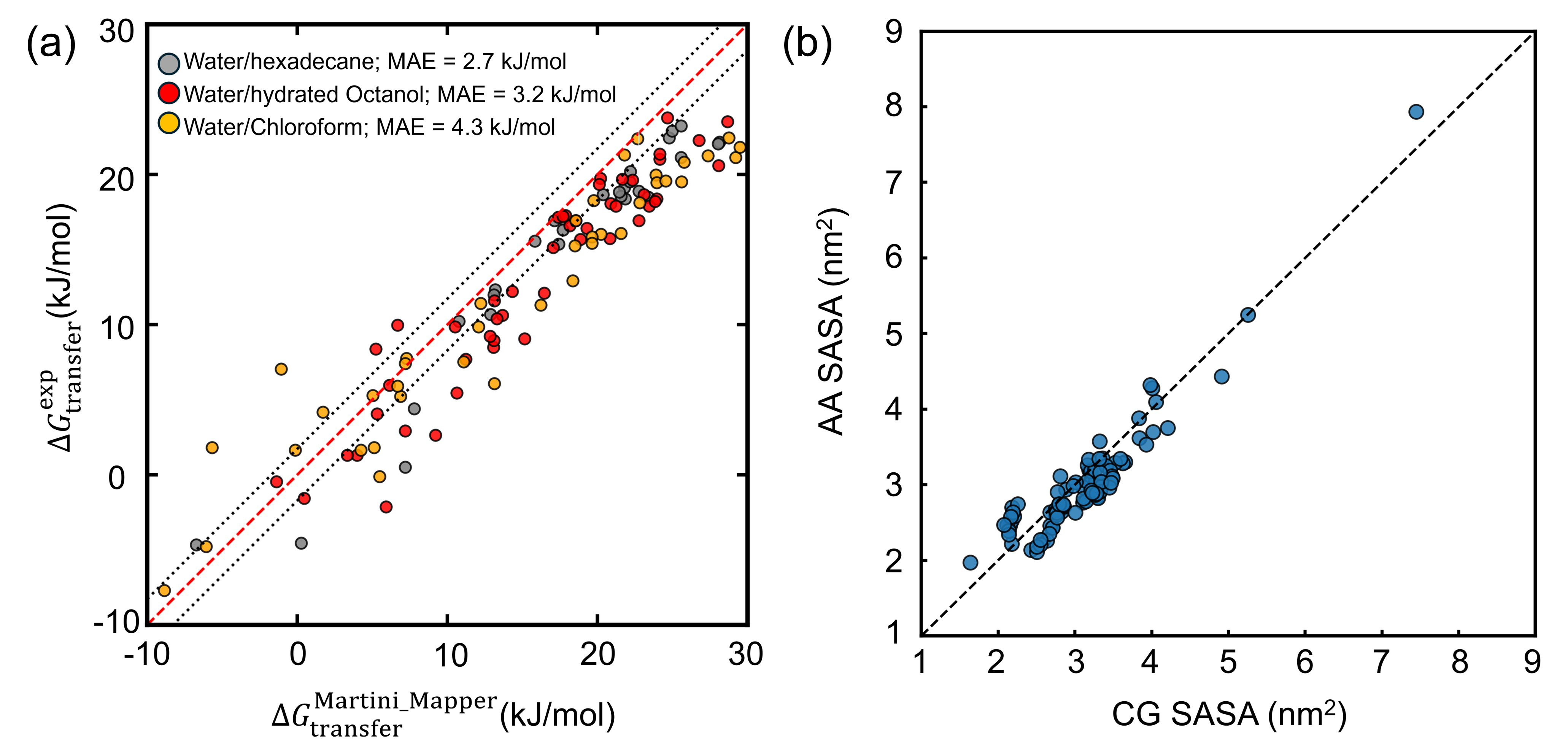}
    \caption{
Thermodynamic and structural validation of Martini\_Mapper on the complete 90-molecule Martini 3 small-molecule dataset. 
(a) Correlation between calculated and experimental transfer free energies ($\Delta G_{\text{transfer}}$) for three solvent pairs: water/hexadecane (gray), water/hydrated octanol (0.2 M water, red), and water/chloroform (yellow). The central dashed line represents ideal agreement ($y = x$), and the dotted lines indicate a $\pm 2.5$ kJ mol$^{-1}$ deviation from the identity line, corresponding to the commonly adopted Martini accuracy threshold. The coefficients of the three types of transfer free energies: water/hexadecane, water/hydrated octanol, and water/chloroform, are 0.82, 0.71, and 0.59, respectively. Experimental data are obtained from \cite{exp1,exp2,exp3,exp4,exp5,exp6}.
(b) Comparison of SASA between CG models generated by Martini\_Mapper and their corresponding AA reference structures from Ref \cite{Alessandri2022}. Each point represents one molecule. The dashed line indicates perfect agreement ($y = x$). The resulting correlation is $R^2 = 0.877$ with RMSE = 0.293 nm$^2$, demonstrating preservation of molecular size and surface exposure in the automated mapping workflow.}
    \label{small90}
\end{figure}

\subsection{Validation on the Martini 3 Small-Molecule Dataset}
\label{small_mol}

Following the protocol of Martini 3 parametrization, we performed here the transfer free energy of three solvent pairs for the Original 90 molecule dataset from the Martini 3 small molecule dataset \cite{Alessandri2022}. A mapping figure of side-by-side comparison of mapped molecules generated by Martini\_Mapper against the human-made model \cite{Alessandri2022} is shown for a set of molecules in Figure S2 of Section S3 in the SI.
The three solvent pairs are the following: water/hydrated octanol (0.2 M water), water/hexadecane, and water/chloroform.
The details of the simulations are given in the Section \ref{simu}.
As mentioned, all transfer free energies were computed using a consistent TI protocol.
The integration timestep for the TI simulations was set to 20 fs or 10 fs, depending on whether the molecule successfully completed the equilibrium stability simulation at the corresponding timestep.
The resultant free energy for a given molecule for each of the solvent pairs is plotted in $x$-axis against the experimental data for that molecule on the $y$-axis.
The experimental data for all the molecules are taken from Ref. \cite{exp1,exp2,exp3,exp4,exp5,exp6}.
We show the correlation between calculated and experimental transfer free energies in Figure~\ref{small90}.

From Figure ~\ref{small90}a, we can see that the correlation yields R$^{2}$ value of 0.82 for water/hexadecane, 0.71 for water/hydrated octanol, and 0.59 for water/chloroform. To further obtain the deviation from the given experimental dataset, we computed the mean absolute error (MAE), which came as 2.7 kJ/mol for water/hexadecane compared to 3.09 kJ/mol obtained by Auto-MartiniM3; 3.2 kJ/mol for water/hydrated octanol compared to 2.33 kJ/mol by Auto-MartiniM3 and 4.3 kJ/mol for water/chloroform compared to 3.57 kJ/mol by Auto-MartiniM3.
It is to be noted that the MAE obtained from Martini\_Mapper is higher than 2.5 kJ/mol, the threshold value adopted in Martini parameterization~\cite{Alessandri2022}; however, the average deviation across all the solvent pairs is not significantly differ than what is obtained using Auto-MartiniM3.
The deviation in Martini\_Mapper is attributed to the absence of virtual site, dihedral, and improper dihedral terms.

In addition to thermodynamic validation, we assessed structural consistency by comparing SASA of the CG models with their corresponding AA reference structures. SASA values were computed using the GROMACS \texttt{gmx sasa} utility with a probe radius of 0.14 nm for AA models and 0.191 nm for CG models, consistent with Martini conventions.\cite{Alessandri2022}
The details of the SASA calculation are detailed in Section S4 of the SI.
As shown in Figure~\ref{small90}b, the CG and AA SASA values exhibit strong correlation, with $R^2 = 0.877$ and RMSE = 0.293 nm$^2$. The CG models show a slight systematic underestimation of SASA, which can be attributed to reduced rigidity in the coarse-grained representation relative to atomistic structures. Nevertheless, the high correlation demonstrates preservation of molecular size and surface exposure across chemically diverse molecules.

All these results confirm that Martini\_Mapper successfully reproduces the Martini 3 small-molecule dataset in a fully automated manner. The workflow generates simulation-ready topologies for all 90 molecules, preserves molecular size and surface characteristics, and reproduces global partitioning trends across multiple solvent environments. While the quantitative accuracy does not yet match manually optimized parametrizations, the framework provides a transparent and reproducible baseline coarse-grained representation that can serve as a foundation for subsequent targeted refinement.

\begin{figure}
    \centering
    \includegraphics[width=1.05\linewidth]{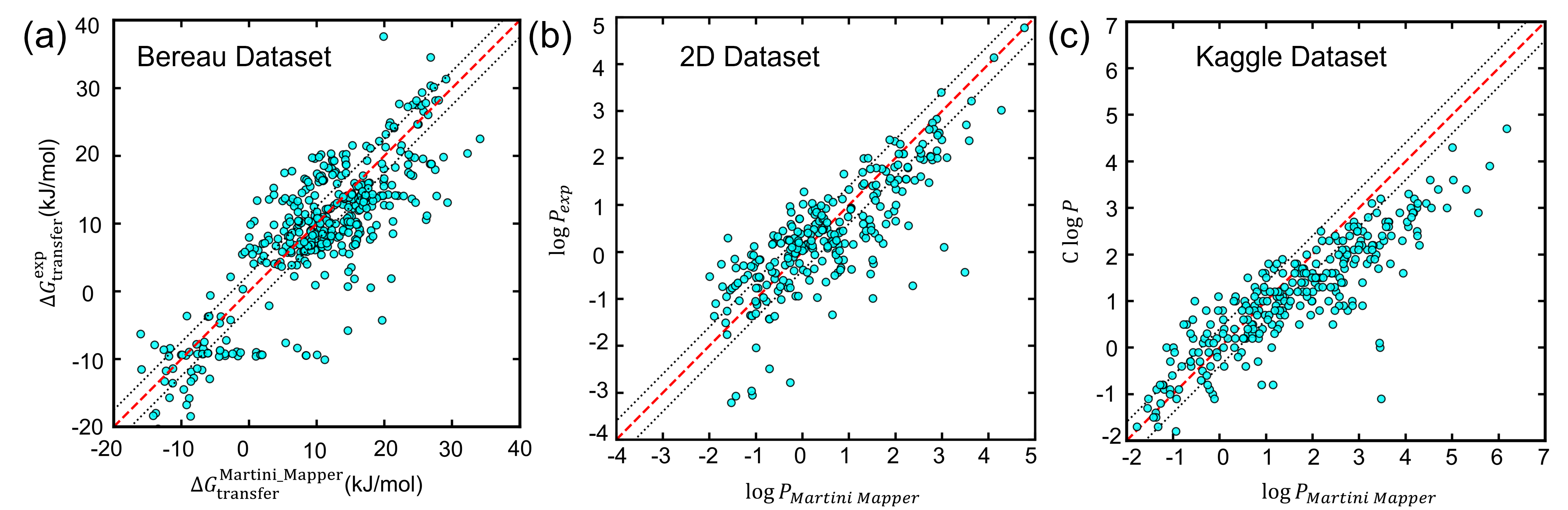}
\caption{Validation of Martini\_Mapper on three independent partitioning datasets: (a) Bereau, (b) 2D dataset, and (c) Kaggle dataset. The dashed line represents ideal agreement ($y = x$), and the dotted lines indicate a $\pm 2.5$ kJ mol$^{-1}$ ($\pm 0.48$ $\log P$ ) deviation from the identity line, corresponding to the commonly adopted Martini accuracy threshold.~\cite{Alessandri2022} $\Delta G_{OW}$ for the Bereau dataset is the transfer free energies from water to hydrated octanol (0.2 M water). For (a) the Bereau dataset, the resulting correlation yields $R^2 = 0.68$, RMSE = 5.69 kJ mol$^{-1}$, and MAE = 4.22 kJ mol$^{-1}$. For (b) the 2D dataset, the resulting metrics are $R^2 = 0.52$, RMSE = 0.838, and MAE = 0.632 in $\log P$ units. For (c) the Kaggle dataset, the resulting performance metrics are $R^2 = 0.40$, RMSE = 0.95, and MAE = 0.72 in $\log P$ units.}
%(a) Correlation between calculated and experimental water–octanol transfer free energies ($\Delta G_{OW}$) for the Bereau dataset. Transfer free energies were computed using Martini\_Mapper-generated coarse-grained topologies without post hoc refinement and compared against experimental reference values. The dashed line represents ideal agreement ($y = x$), and the dotted lines indicate a $\pm 2.5$ kJ mol$^{-1}$ deviation from the identity line, corresponding to the commonly adopted Martini accuracy threshold. The resulting correlation yields $R^2 = 0.68$, RMSE = 5.69 kJ mol$^{-1}$, and MAE = 4.22 kJ mol$^{-1}$.
%(b) Correlation between predicted and experimental $\log P$ values for the 2D benchmark dataset. The dashed line represents perfect agreement, and the dotted lines indicate a $\pm 0.48$ $\log P$ deviation (equivalent to $\pm 2.5$ kJ mol$^{-1}$). The resulting metrics are $R^2 = 0.52$, RMSE = 0.838, and MAE = 0.632 in $\log P$ units.
%(c) Correlation between predicted and experimental $\log P$ values for the Kaggle dataset. The dashed line denotes ideal agreement, and the dotted lines represent a $\pm 0.48$ $\log P$ deviation from the identity line. The resulting performance metrics are $R^2 = 0.40$, RMSE = 0.95, and MAE = 0.72 in $\log P$ units.}}
\label{validation_independent}
\end{figure}

\subsection{Validation on Independent Partitioning Datasets}

To evaluate the predictive reliability of Martini\_Mapper as a fully automated coarse-graining framework, we assessed its performance on three independent datasets of experimentally measured water--octanol partitioning thermodynamics. All molecules were mapped using the LBBT dictionary in a single-pass automated workflow, and transfer free energies were computed using a consistent simulation protocol across datasets. It is to note that only molecules that successfully completed 10 ns NPT stability simulations in pure water (as described in Section \ref{simu}) were subsequently subjected to TI calculations using the 20 fs timestep. All the itp/gro files of these datasets are available in GitHub:
\url{https://github.com/eliobaby/Martini_Mapper/tree/main/Working_Molecules}.

\paragraph{Bereau Dataset:} The TI simulation results for the Bereau dataset include 427 molecules that were successfully completed under NPT conditions. This dataset consists of structurally diverse neutral organic compounds with experimentally measured water–octanol transfer free energies. Figure~\ref{validation_independent}a shows the correlation between calculated and experimental $\Delta G_{OW}$ values for the working subset. The resulting correlation yields $R^2 = 0.68$, RMSE = 5.69 kJ mol$^{-1}$, and MAE = 4.22 kJ mol$^{-1}$. 
%Using the Martini accuracy threshold of $\pm 2.5$ kJ mol$^{-1}$, 244 molecules exceed this deviation, and 177 exceed $\pm 4.0$ kJ mol$^{-1}$. 
While systematic deviations remain, the overall trend in hydrophobicity is captured across a broad chemical space without molecule-specific tuning. These results indicate that Martini\_Mapper provides a consistent baseline description of partitioning thermodynamics, suitable as an initial automated model that can subsequently be refined when higher quantitative accuracy is required.

\paragraph{2D Benchmark Dataset:} We next evaluated generalizability using the independent 2D molecular benchmark dataset. In this dataset, we ran TI of 267 molecules, which were successfully run with 20 fs NPT production run for 10 ns. As shown in Figure~\ref{validation_independent}b, the correlation between predicted and experimental $\log P$ values yields $R^2 = 0.52$, RMSE = 0.838, and MAE = 0.632 in $\log P$ units. 
%A total of 145 molecules exceed a $\pm 0.45$ $\log P$ deviation (equivalent to $\pm 2.5$ kJ mol$^{-1}$), and 103 exceed $\pm 0.68$ $\log P$ (approximately $\pm 4.0$ kJ mol$^{-1}$). 
The preservation of moderate correlation on an unseen dataset indicates that the workflow captures global physicochemical trends despite the absence of dataset-specific refinement.

\paragraph{Kaggle Dataset}: Finally, we examined performance on the Kaggle logP dataset, which contains chemically diverse small molecules. In this dataset, we ran TI of 291 molecules, which were successfully run with 20 fs NPT production run for 10 ns. The resulting correlation (Figure~\ref{validation_independent}c) yields $R^2 = 0.40$, RMSE = 0.95, and MAE = 0.72 in $\log P$ units. 
%In this dataset, 178 molecules exceed $\pm 0.45$ $\log P$, and 128 exceed $\pm 0.68$ $\log P$. 
The reduced correlation relative to the Bereau and 2D datasets is reasonable with broader chemical diversity and fragment coverage limitations of the current dictionary, which affect the quantitative fidelity of the first-pass CG representation.

Across all three independent datasets, the results demonstrate that Martini\_Mapper provides a consistent and fully automated baseline coarse-grained representation without molecule-specific optimization or iterative tuning. The observed quantitative deviations therefore reflect the combined effects of dictionary coverage, bead assignment choices, bonded-term simplifications, and the absence of dataset-specific refinement. Rather than attributing these deviations to intrinsic limitations of coarse-graining, the present benchmarks define the current performance envelope of the automated workflow under a fixed and reproducible protocol.
%Future improvements in dictionary expansion, bonded parametrization (including dihedrals where necessary), and systematic refinement strategies are expected to improve quantitative agreement while maintaining automation and scalability.}

\subsection{Structural Validation of Larger Molecules}

\begin{figure}
\centering
\includegraphics[width=0.7\linewidth]{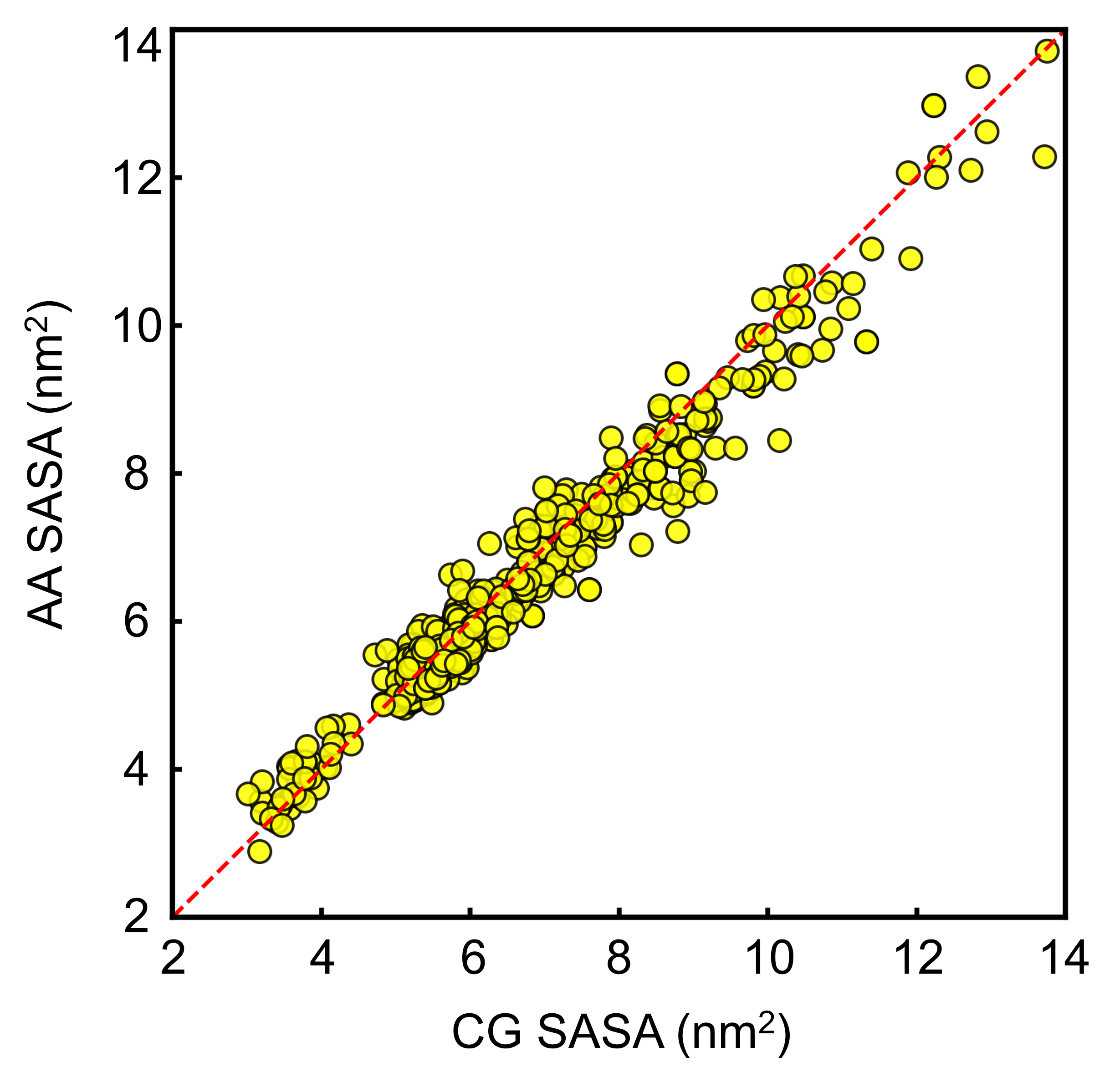}
\caption{Comparison of SASA for 560 TPCN molecules between coarse-grained models generated by Martini\_Mapper and corresponding all-atom reference structures obtained from the xTB pipeline. Each point represents one molecule. The dashed red line indicates ideal agreement ($y=x$). The correlation between CG and AA SASA values is $R^2 = 0.960$ with RMSE = 0.401~nm$^2$, demonstrating strong preservation of molecular volume and surface characteristics across systems containing 9-75 heavy atoms.}
\label{sasa_tpcn}
\end{figure}

To evaluate the structural consistency of coarse-grained models for larger systems, we performed a quantitative comparison of SASA between Martini\_Mapper-generated CG models and their corresponding AA reference structures of molecules from the TPCN dataset. The AA structures were obtained from our xTB-based optimization pipeline, ensuring internal methodological consistency.
Due to the large number of constraints present in highly rigid systems and the current absence of automated virtual-site generation, full thermodynamic validation via transfer free energy calculations was not performed for the entire TPCN dataset. Instead, SASA was used as a structural validation metric, consistent with established Martini 3 parametrization principles where preservation of molecular volume and surface characteristics is essential.
A representative subset of 560 molecules from the TPCN dataset was analyzed, spanning systems containing 9 to 75 heavy atoms. The path to this dataset is GitHub:
\url{https://github.com/eliobaby/Martini_Mapper/tree/main/Working_Molecules/TPCN/xtb_2}. As shown in Figure~\ref{sasa_tpcn}, CG and AA SASA values exhibit excellent agreement with a correlation coefficient of $R^2 = 0.960$ and RMSE = 0.401~nm$^2$. The strong linear correlation indicates that the mapping procedure preserves overall molecular size and surface properties even for larger and more complex systems. It is to be noted that the full set of 4,716 molecules mapped using Martini\_Mapper was not included in the SASA comparison to reduce computational cost. This exclusion was solely due to resource considerations and does not indicate any limitations of the Martini\_Mapper.

\subsection{Computational Performance and Numerical Stability}
\label{success}

\begin{figure}[htbp]
\centering
\includegraphics[width=\textwidth]{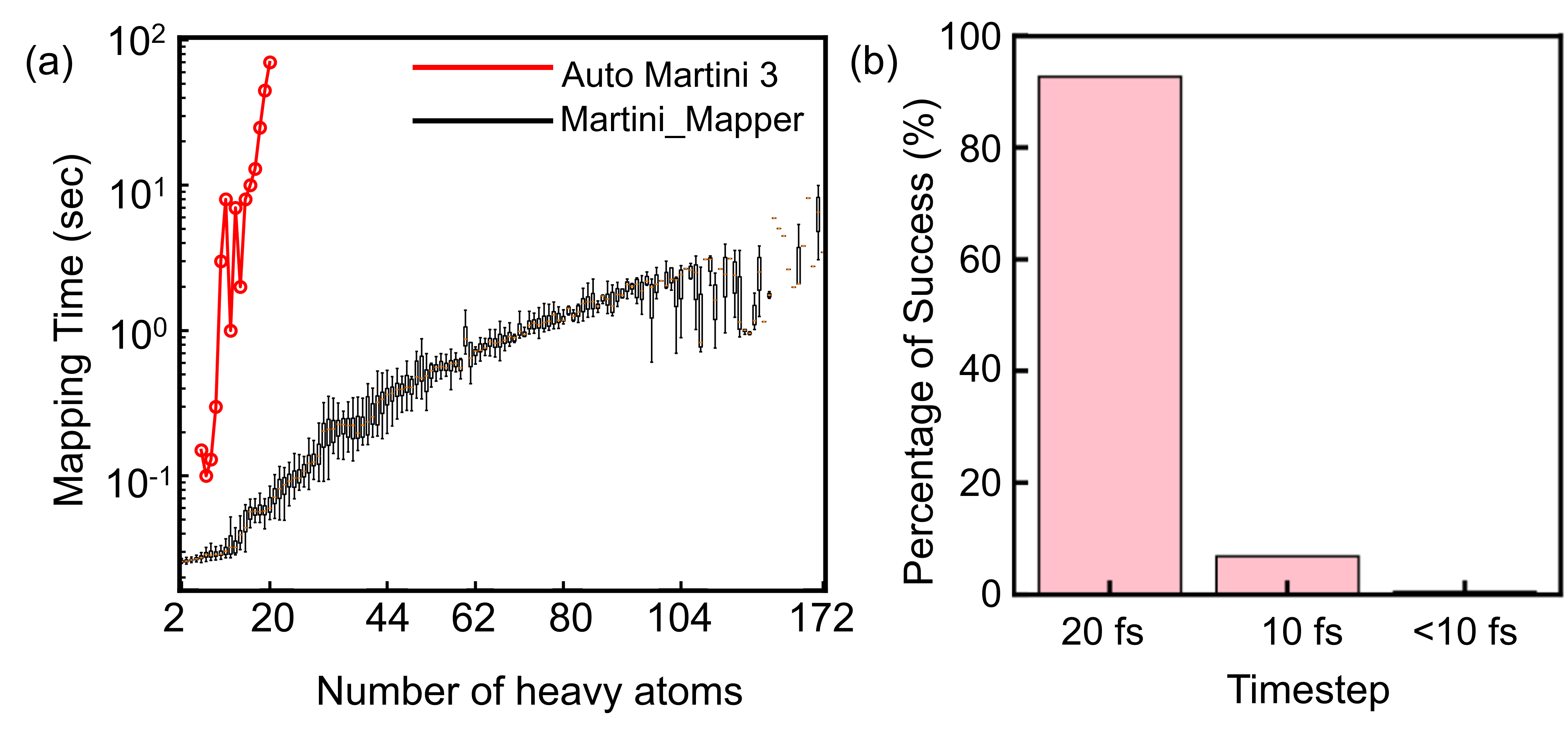}
\caption{
Performance and stability analysis of the Martini\_Mapper workflow. 
(a) Core mapping time as a function of molecular size (number of heavy atoms). Each data point corresponds to a single molecule mapped using one CPU core (Intel Xeon E5-2680v2 processor). The dataset includes molecules from the original 90-molecule benchmark, Bereau set, 2D dataset, Kaggle dataset, Gr{\"u}newald set, and the TPCN database. Black symbols represent Martini\_Mapper timings obtained without the bond/angle potential parameter refinement stage, while red symbols indicate mapping times obtained using Auto-MartiniM3~\cite{Szczuka2025}. The near-linear trend on the log–log scale demonstrates scalable performance of the Martini\_Mapper framework. 
(b) Percentage of successful coarse-grained molecular dynamics simulations as a function of integration timestep. The majority ($>90$\%) of automatically generated models remain stable at the standard Martini timestep of 20 fs, indicating that the automatically generated bonded parameters yield numerically stable models for standard Martini simulation settings. In testing the success percentage, all the simulations were performed in NPT ensembles with a production runtime of 10 ns.
}
\label{performance}
\end{figure}

In addition to thermodynamic and structural validation, we evaluated the computational efficiency and scalability of the Martini\_Mapper workflow. The core mapping time as a function of molecular size (number of heavy atoms) is presented in Figure~\ref{performance}. The reported timings correspond exclusively to the rule-based mapping stage and explicitly exclude the xTB-based coordinate generation and bonded parameter refinement step (i.e., executed using the \texttt{--no-xtb} flag). Each molecule was mapped using a single CPU core (Intel Xeon E5-2680v2 processor). The results demonstrate near-linear scaling of mapping time with molecular size across a broad range of compounds, indicating that the algorithm maintains computational tractability even for larger fragments.

For example, a molecule containing 20 heavy atoms requires approximately 0.07 s using Martini\_Mapper, whereas Auto-MartiniM3\cite{Szczuka2025} requires approximately 70 s under comparable single-molecule conditions. This difference arises from the fully deterministic and fragment-based nature of the Martini\_Mapper algorithm. The observed scaling behavior confirms that the computational complexity is dominated by graph traversal and fragment assignment operations, which scale approximately linearly with molecular size. Consequently, the automated mapping procedure achieves substantially lower wall-time per molecule while remaining systematic and reproducible. These timings correspond exclusively to the core fragment-based mapping stage (executed with the \texttt{--no-xtb} flag). The difference reflects the deterministic fragment-based mapping strategy employed in Martini\_Mapper.

We further assessed the numerical robustness of the generated coarse-grained models by performing CG molecular dynamics simulations using different integration timesteps. The percentage of successful simulations as a function of timestep is also shown in Figure~\ref{performance}. The majority of automatically generated models remain stable at the standard Martini timestep of 20 fs, demonstrating that the bonded parameters and constraint treatment produce numerically stable systems suitable for routine CG-MD simulations. These results confirm that the workflow combines computational efficiency with physically consistent model generation. These benchmarks provide a comprehensive evaluation framework encompassing thermodynamic accuracy, structural consistency, computational efficiency, and simulation robustness, enabling direct comparison of Martini\_Mapper with existing automated and manually curated Martini 3 parametrization approaches.\cite{Szczuka2025,Alessandri2022}

\section{Limitations of Martini\_Mapper}

While the automated framework substantially advances the process of generating Martini~3 coarse-grained models, several important limitations remain. The first limitation we want to discuss is related to our bead dictionary. As a result, the current ruleset is most reliable for molecules dominated by carbon (C), oxygen (O), and nitrogen (N), and less robust when encountering sulfur-, phosphorus-, or halogen-containing fragments, or other motifs not present in the original dataset. 
Therefore, it is fundamentally more successful in mapping, which is tied to the coverage of its bead dictionary.
%To explain this limit in detail, for the Bereau training set, 542 out of 653 molecules (87.9\%) could be mapped without error, whereas for the independent 2D, Kaggle, and TPCN test sets the overall mappability (including DBBT), ranging from about 37\% for the chemically diverse Kaggle set to 75\% for the 2D benchmark.  These numbers explain both the strength of the general algorithm and the need to expand the dictionary with additional rules to achieve more wider coverage.

The mapping procedure itself introduces algorithmic constraints. The strict requirement that path lengths within a bead remain three bonds or fewer ensures physical plausibility, but it can also exclude chemically reasonable mappings for highly branched structures or unusual topologies. Enforcing molecular symmetry can fail when stereocenters or asymmetric substituents are present, since chirality is not explicitly represented. Recursive splitting of large fragments may yield multiple mathematically valid partitions, but the current algorithm lacks a mechanism to rank these alternatives by chemical realism. These features reflect the deterministic and rule-based nature of the present implementation.

%Although we assume the error contribution of each bead type is linearly combined, contributing to the final deviation of log P from the experimental data, we do not exclude other nonlinear combinations, and even adapt an advanced optimization algorithm to refine the bead type. The accuracy of the final bead assignments is also conditioned by the limits of Martini~3 itself. Although Martini~3 broadened its bead vocabulary and introduced finer resolution compared to earlier versions, the vocabulary remains finite and does not directly capture all chemical environments. In cases where no native bead exists, the algorithm must enforce the ``closest match,” introducing systematic biases. Resolution is inherently limited to three or four heavy atoms per bead, which means stereochemistry, chirality, and subtle conformational preferences are lost, even when the mapping is otherwise optimal. Furthermore, Martini parameterization emphasizes broad thermodynamic trends, such as partition coefficients or solvation free energies, and is not designed to reproduce subtle, environment-dependent interactions or strongly directional bonding. The increased chemical specificity of Martini~3, while a strength, also raises the bar for automated mapping: bead choices are more context-sensitive, and incomplete dictionary coverage becomes a more significant bottleneck.

The present implementation performs a single-pass bead assignment without iterative optimization against multiple thermodynamic targets. Bead types are selected deterministically from the dictionary and are not refined through feedback from experimental observables. As a result, the reported accuracy represents a reproducible first-pass parametrization rather than a fully optimized Martini 3 model.
Furthermore, though the current workflow automatically generates bond and angle parameters, it does not generate proper or improper dihedral terms. Planarity in rigid or fused aromatic systems is therefore maintained primarily through the angle network and constraints. Automated generation of dihedral and improper potentials remains a limitation of the present framework.
Similarly, virtual-site construction is not implemented. Manually curated Martini models often introduce virtual sites to preserve rigidity and improve numerical stability in planar systems. The absence of automated virtual-site generation may limit structural fidelity in certain highly rigid molecules.
Very stiff bonded interactions are treated thus as constraints to maintain numerical stability at standard Martini timesteps. While this approach ensures robust simulations, it may restrict fine control over highly rigid or torsionally complex systems.

Finally, validation across independent datasets reflects the current performance envelope of the automated workflow under a fixed and reproducible protocol. The observed deviations across chemically diverse datasets therefore reflect the combined effects of dictionary coverage, deterministic bead assignment, bonded-term simplifications, and the absence of dataset-specific refinement. Future development will focus on expanding dictionary coverage, incorporating automated dihedral and virtual-site generation, and introducing iterative optimization strategies to improve transferability while preserving automation.

%Finally, validation of the generated models reflects these combined challenges. Bead parameters tuned against a moderately sized dataset of 653 molecules (Bereau) achieved strong agreement with experiment, yielding an $R^2$ score of 0.83 for \(\Delta G_{OW}\) predictions by testing 481 molecules. However, when applied to independent test sets, the correlation dropped to 0.71 for the Kaggle dataset and 0.63 for the 2D dataset in predicting log$P$, indicating bias towards the training set and limited transferability across chemical space. This performance is respectable for a coarse-grained framework, but also emphasizes the importance of extending both the dictionary and the calibration data. Another important limitation is that our present framework lacks automated rules for assigning charged beads. This restricts the mapping of ionizable groups, which are frequently encountered in polyzwitterions.
%Together, these constraints reflect a mixture of algorithmic boundaries and fundamental limitations of Martini~3, which must be addressed through dictionary expansion, more diverse calibration, and potentially hybrid integration with data-driven or machine-learning approaches in the future.
%We are also extending the method to include an automated optimization step, allowing alternative bead assignments to be compared and refined in a systematic way.

\section{Conclusion}

In this work, we introduced a fully automated framework that transforms  SMILES strings of a molecule into Martini~3 coarse-grained models through a systematic, rule-based algorithm.
By formalizing the mapping procedure into a dictionary-driven workflow, the approach reduces subjective variability in bead assignment and reliance on chemical intuition, thereby addressing a practical bottleneck in high-throughput coarse-grained model construction.
The framework was evaluated on chemically diverse datasets and benchmarked against experimental $\log P$ and transfer free energy values across multiple solvent systems. The benchmarking results demonstrate that automated mapping can produce reproducible, simulation-ready Martini~3 topologies across molecules ranging from small systems to compounds up to 172 heavy atoms, with quantitative performance comparable to existing automated approaches. In total, our framework successfully mapped 6,280 molecules across six datasets: 542 from Bereau, 300 from 2D, 332 from Gr{\"u}newald, 300 from Kaggle, 4,716 from TPCN, and 90 from the Original 90 molecule dataset.

The significance of this advance lies in making coarse-grained simulation more accessible, reproducible, and scalable. The primary challenge in small-molecule parametrization is not the absence of standards, but the consistent application of detailed guidelines in high-throughput contexts. Martini\_Mapper is designed to automate and scale the systematic application of established Martini 3 conventions.
By replacing manual mapping with an automated and extensible pipeline, the framework enables systematic treatment of large chemical libraries while maintaining consistency across molecules.
In direct comparison with recent automated methodologies such as AutoMartini3, the present implementation achieves thermodynamic accuracy comparable to recent automated approaches, while substantially reducing computational overhead during the core mapping stage.

While the current framework establishes a robust foundation, several clear avenues remain for future development. The most immediate priority is the systematic expansion of the bead dictionary. Although the present rules capture a broad range of carbon-, oxygen-, and nitrogen-containing chemistries, coverage of sulfur, phosphorus, halogens, and metal coordination environments remains incomplete. 
Targeted inclusion of fragments from chemically diverse libraries will progressively reduce mapping failures and improve coverage across broader chemical space.
A second direction concerns refinement of bonded and nonbonded parameter treatment. The present workflow performs deterministic bead assignment without iterative refinement against multiple thermodynamic targets.
Systematic expansion of the benchmarking dataset, and incorporation of automated dihedral and virtual-site generation would improve transferability and structural fidelity.
Validation against observables beyond $\log P$ and $\Delta G_{OW}$ may further strengthen robustness across chemically diverse systems.

In conclusion, these results present the current implementation of Martin\_Mapper as a reproducible and scalable baseline for automated Martini~3 parametrization. Continued development will focus on expanding chemical coverage, incorporating higher-order bonded terms, and improving transferability through systematic validation, while maintaining the deterministic and high-throughput character of the workflow.

\section*{Supporting Information}

The Supporting Information includes additional materials that complement the main manuscript. \textbf{Section S1} presents the formal derivation of bond and angle equilibrium values and corresponding force constants. \textbf{Section S2} and \textbf{Figure S1} report hydration free-energy benchmarks for 106 Martini 3 beads, comparing thermodynamic integration results obtained in this work with reference Martini 3 values. \textbf{Section S3} and \textbf{Figure S2} compares representative small-molecule mappings generated by Martini\_Mapper with expert-curated Martini 3 models. \textbf{Section S4} details the SASA calculation protocol for both atomistic and coarse-grained models.

\section*{Data and Software Availability}
The data underlying this study, including all the gro/itp files of the working molecules and open source codes for Martini Mapper, are openly available at 
\url{https://github.com/eliobaby/Martini_mapper}.

\section*{Author Contributions}
K.V.B. developed the algorithm, performed data analysis, prepared the figures, and wrote and revised the initial draft of the manuscript. S.N. tested the dataset, carried out additional analyses, prepared figures, and wrote and revised the full manuscript. Y.A. conceived and conceptualized the study, supervised the research, contributed to data interpretation, reviewed the manuscript, and acquired funding. We sincerely thank the anonymous reviewers for their valuable feedback and insightful suggestions, which have helped improve the quality of this manuscript. All authors discussed the results and approved the manuscript.

\section*{Conflicts of interest}
There are no conflicts of interest to declare.

\section*{Acknowledgements}
An, Y. acknowledges the support from the Louisiana Board of Regents  RSC funding (RA-D-05), National Science
Foundation (OIA-1946231) and the Louisiana Board of Regents for the
Louisiana Materials Design Alliance (LAMDA) and the LSU start-up fund. The simulations were carried out on the LSU-HPC facilities and the LONI HPC facilities.

\bibliography{ref}

@incollection{Frenkel2002,
booktitle = {Understanding Molecular Simulation (Second Edition)},
publisher = {Academic Press},
edition = {Second Edition},
address = {San Diego},
year = {2002},
isbn = {978-0-12-267351-1},
author = {Daan Frenkel and Berend Smit}
}

@incollection{Leach2002,
author = {Andrew Leach},
booktitle = {Molecular Modeling: Principles and Applications},
publisher = {Pearson Education},
edition = {Second Edition},
year = {2002}
}

@Article{Karplus2002,
author={Karplus, Martin
and McCammon, J. Andrew},
title={Molecular dynamics simulations of biomolecules},
journal={Nature Structural Biology},
year={2002},
month={Sep},
day={01},
volume={9},
number={9},
pages={646-652},
}

@article{Hollingsworth2012,
title = {Molecular Dynamics Simulation for All},
journal = {Neuron},
volume = {99},
number = {6},
pages = {1129-1143},
year = {2018},
issn = {0896-6273},
author = {Scott A. Hollingsworth and Ron O. Dror},
}

@article{Shaw2010,
author = {David E. Shaw  and Paul Maragakis  and Kresten Lindorff-Larsen  and Stefano Piana  and Ron O. Dror  and Michael P. Eastwood  and Joseph A. Bank  and John M. Jumper  and John K. Salmon  and Yibing Shan  and Willy Wriggers },
title = {Atomic-Level Characterization of the Structural Dynamics of Proteins},
journal = {Science},
volume = {330},
number = {6002},
pages = {341-346},
year = {2010},
}

@Article{Meller2023,
author={Meller, Artur
and Bhakat, Soumendranath
and Solieva, Shahlo
and Bowman, Gregory R.},
title={Accelerating Cryptic Pocket Discovery Using AlphaFold},
journal={Journal of Chemical Theory and Computation},
year={2023},
month={Jul},
day={25},
publisher={American Chemical Society},
volume={19},
number={14},
pages={4355-4363},
}

@Article{Zuzic2022,
author={Zuzic, Lorena
and Samsudin, Firdaus
and Shivgan, Aishwary T.
and Raghuvamsi, Palur V.
and Marzinek, Jan K.
and Boags, Alister
and Pedebos, Conrado
and Tulsian, Nikhil K.
and Warwicker, Jim
and MacAry, Paul
and Crispin, Max
and Khalid, Syma
and Anand, Ganesh S.
and Bond, Peter J.},
title={Uncovering cryptic pockets in the SARS-CoV-2 spike glycoprotein},
journal={Structure},
year={2022},
month={Aug},
day={04},
volume={30},
number={8},
pages={1062-1074.e4},
}

@Inbook{Zhou2022,
author={Zhou, Kun
and Liu, Bo},
title={Preface},
bookTitle={Molecular Dynamics Simulation},
year={2022},
month={Jan},
day={01},
publisher={Elsevier},
}

@Book{Rapaport2004,
author={Rapaport, D. C.},
title={The Art of Molecular Dynamics Simulation},
year={2004},
edition={2},
publisher={Cambridge University Press},
address={Cambridge},
}

@Article{DeVivo2016,
author={De Vivo, Marco
and Masetti, Matteo
and Bottegoni, Giovanni
and Cavalli, Andrea},
title={Role of Molecular Dynamics and Related Methods in Drug Discovery},
journal={Journal of Medicinal Chemistry},
year={2016},
month={May},
day={12},
publisher={American Chemical Society},
volume={59},
number={9},
pages={4035-4061},
issn={0022-2623},
}

@Article{Cornell1995,
author={Cornell, Wendy D.
and Cieplak, Piotr
and Bayly, Christopher I.
and Gould, Ian R.
and Merz, Kenneth M.
and Ferguson, David M.
and Spellmeyer, David C.
and Fox, Thomas
and Caldwell, James W.
and Kollman, Peter A.},
title={A Second Generation Force Field for the Simulation of Proteins, Nucleic Acids, and Organic Molecules},
journal={Journal of the American Chemical Society},
year={1995},
month={May},
day={01},
publisher={American Chemical Society},
volume={117},
number={19},
pages={5179-5197},
issn={0002-7863},
}

@article{Noid2013,
    author = {Noid, W. G.},
    title = {Perspective: Coarse-grained models for biomolecular systems},
    journal = {The Journal of Chemical Physics},
    volume = {139},
    number = {9},
    pages = {090901},
    year = {2013},
    month = {09},
}

@article{Klein2008,
author = {Michael L. Klein  and Wataru Shinoda },
title = {Large-Scale Molecular Dynamics Simulations of Self-Assembling Systems},
journal = {Science},
volume = {321},
number = {5890},
pages = {798-800},
year = {2008},
}

@Article{Kmiecik2016,
author={Kmiecik, Sebastian
and Gront, Dominik
and Kolinski, Michal
and Wieteska, Lukasz
and Dawid, Aleksandra Elzbieta
and Kolinski, Andrzej},
title={Coarse-Grained Protein Models and Their Applications},
journal={Chemical Reviews},
year={2016},
month={Jul},
day={27},
publisher={American Chemical Society},
volume={116},
number={14},
pages={7898-7936},
}

@Article{Marrink2004,
author={Marrink, Siewert J.
and de Vries, Alex H.
and Mark, Alan E.},
title={Coarse Grained Model for Semiquantitative Lipid Simulations},
journal={The Journal of Physical Chemistry B},
year={2004},
month={Jan},
day={01},
publisher={American Chemical Society},
volume={108},
number={2},
pages={750-760},
issn={1520-6106},
}

@Article{Souza2021,
author={Souza, Paulo C. T.
and Alessandri, Riccardo
and Barnoud, Jonathan
and Thallmair, Sebastian
and Faustino, Ignacio
and Gr{\"u}newald, Fabian
and Patmanidis, Ilias
and Abdizadeh, Haleh
and Bruininks, Bart M. H.
and Wassenaar, Tsjerk A.
and Kroon, Peter C.
and Melcr, Josef
and Nieto, Vincent
and Corradi, Valentina
and Khan, Hanif M.
and Doma{\'{n}}ski, Jan
and Javanainen, Matti
and Martinez-Seara, Hector
and Reuter, Nathalie
and Best, Robert B.
and Vattulainen, Ilpo
and Monticelli, Luca
and Periole, Xavier
and Tieleman, D. Peter
and de Vries, Alex H.
and Marrink, Siewert J.},
title={Martini 3: a general purpose force field for coarse-grained molecular dynamics},
journal={Nature Methods},
year={2021},
month={Apr},
day={01},
volume={18},
number={4},
pages={382-388},
}

@article{Alessandri2022,
author = {Alessandri, Riccardo and Barnoud, Jonathan and Gertsen, Anders S. and Patmanidis, Ilias and de Vries, Alex H. and Souza, Paulo C. T. and Marrink, Siewert J.},
title = {Martini 3 Coarse-Grained Force Field: Small Molecules},
journal = {Advanced Theory and Simulations},
volume = {5},
number = {1},
pages = {2100391},
year = {2022}
}

@Article{Bereau2015,
author={Bereau, Tristan
and Kremer, Kurt},
title={Automated Parametrization of the Coarse-Grained Martini Force Field for Small Organic Molecules},
journal={Journal of Chemical Theory and Computation},
year={2015},
month={Jun},
day={09},
publisher={American Chemical Society},
volume={11},
number={6},
pages={2783-2791},
issn={1549-9618},
doi={10.1021/acs.jctc.5b00056},
url={https://doi.org/10.1021/acs.jctc.5b00056}
}

@Article{Husic2020,
author={Husic, Brooke E.
and Charron, Nicholas E.
and Lemm, Dominik
and Wang, Jiang
and P{\'e}rez, Adri{\`a}
and Majewski, Maciej
and Kr{\"a}mer, Andreas
and Chen, Yaoyi
and Olsson, Simon
and de Fabritiis, Gianni
and No{\'e}, Frank
and Clementi, Cecilia},
title={Coarse graining molecular dynamics with graph neural networks},
journal={The Journal of Chemical Physics},
year={2020},
month={Nov},
day={16},
volume={153},
number={19},
pages={194101},
}

@article{Bolhuis2024,
    author = {del Razo, Mauricio J. and Crommelin, Daan and Bolhuis, Peter G.},
    title = {Data-driven dynamical coarse-graining for condensed matter systems},
    journal = {The Journal of Chemical Physics},
    volume = {160},
    number = {2},
    pages = {024108},
    year = {2024},
    month = {01},
}

@inproceedings{Niki2022,
author = {Nasikas, Dimitris and Ricci, Eleonora and Giannakopoulos, George and Karkaletsis, Vangelis and Theodorou, Doros N. and Vergadou, Niki},
title = {Investigation of Machine Learning-based Coarse-Grained Mapping Schemes for Organic Molecules},
year = {2022},
isbn = {9781450395977},
publisher = {Association for Computing Machinery},
booktitle = {Proceedings of the 12th Hellenic Conference on Artificial Intelligence},
articleno = {51},
}

@article{Kroon_2025,
title={Martinize2 and Vermouth: Unified Framework for Topology Generation},
publisher={eLife Sciences Publications, Ltd},
author={Kroon, Peter C and Gr{\"u}newald, Fabian and Barnoud, Jonathan and van Tilburg, Marco and Brasnett, Chris and de Souza, Paulo Cesar Telles and Wassenaar, Tsjerk A and Marrink, Siewert-Jan J},
year={2025}, 
}

@Article{Potter2021,
author={Potter, Thomas D.
and Barrett, Elin L.
and Miller, Mark A.},
title={Automated Coarse-Grained Mapping Algorithm for the Martini Force Field and Benchmarks for Membrane--Water Partitioning},
journal={Journal of Chemical Theory and Computation},
year={2021},
month={Sep},
day={14},
publisher={American Chemical Society},
volume={17},
number={9},
pages={5777-5791},
}

@Article{Grunewald2022,
author={Gr{\"u}newald, Fabian
and Punt, Mats H.
and Jefferys, Elizabeth E.
and Vainikka, Petteri A.
and K{\"o}nig, Melanie
and Virtanen, Valtteri
and Meyer, Travis A.
and Pezeshkian, Weria
and Gormley, Adam J.
and Karonen, Maarit
and Sansom, Mark S. P.
and Souza, Paulo C. T.
and Marrink, Siewert J.},
title={Martini 3 Coarse-Grained Force Field for Carbohydrates},
journal={Journal of Chemical Theory and Computation},
year={2022},
month={Dec},
day={13},
publisher={American Chemical Society},
volume={18},
number={12},
pages={7555-7569},
}

@Article{Alessandri2019,
author={Alessandri, Riccardo
and Souza, Paulo C. T.
and Thallmair, Sebastian
and Melo, Manuel N.
and de Vries, Alex H.
and Marrink, Siewert J.},
title={Pitfalls of the Martini Model},
journal={Journal of Chemical Theory and Computation},
year={2019},
month={Oct},
day={08},
publisher={American Chemical Society},
volume={15},
number={10},
pages={5448-5460},
}

@article{Voth2013,
   author = "Saunders, Marissa G. and Voth, Gregory A.",
   title = "Coarse-Graining Methods for Computational Biology", 
   journal= "Annual Review of Biophysics",
   year = "2013",
   volume = "42",
   number = "Volume 42, 2013",
   pages = "73-93",
   doi = "https://doi.org/10.1146/annurev-biophys-083012-130348",
   url = "https://www.annualreviews.org/content/journals/10.1146/annurev-biophys-083012-130348",
   publisher = "Annual Reviews",
   issn = "1936-1238",
  }

@Article{Monticelli2008,
author={Monticelli, Luca
and Kandasamy, Senthil K.
and Periole, Xavier
and Larson, Ronald G.
and Tieleman, D. Peter
and Marrink, Siewert-Jan},
title={The MARTINI Coarse-Grained Force Field: Extension to Proteins},
journal={Journal of Chemical Theory and Computation},
year={2008},
month={May},
day={01},
publisher={American Chemical Society},
volume={4},
number={5},
pages={819-834},
}

@Article{Marrink2007,
author={Marrink, Siewert J.
and Risselada, H. Jelger
and Yefimov, Serge
and Tieleman, D. Peter
and de Vries, Alex H.},
title={The MARTINI Force Field:{\thinspace} Coarse Grained Model for Biomolecular Simulations},
journal={The Journal of Physical Chemistry B},
year={2007},
month={Jul},
day={01},
publisher={American Chemical Society},
volume={111},
number={27},
pages={7812-7824}
}

@Article{MacKerell1998,
author={MacKerell Jr., A. D.
and Bashford, D.
and Bellott, M.
and Dunbrack Jr., R. L.
and Evanseck, J. D.
and Field, M. J.
and Fischer, S.
and Gao, J.
and Guo, H.
and Ha, S.
and Joseph-McCarthy, D.
and Kuchnir, L.
and Kuczera, K.
and Lau, F. T. K.
and Mattos, C.
and Michnick, S.
and Ngo, T.
and Nguyen, D. T.
and Prodhom, B.
and Reiher, W. E.
and Roux, B.
and Schlenkrich, M.
and Smith, J. C.
and Stote, R.
and Straub, J.
and Watanabe, M.
and Wi{\'o}rkiewicz-Kuczera, J.
and Yin, D.
and Karplus, M.},
title={All-Atom Empirical Potential for Molecular Modeling and Dynamics Studies of Proteins},
journal={The Journal of Physical Chemistry B},
year={1998},
month={Apr},
day={01},
publisher={American Chemical Society},
volume={102},
number={18},
pages={3586-3616},
}

@Article{Casalino2020,
author={Casalino, Lorenzo
and Gaieb, Zied
and Goldsmith, Jory A.
and Hjorth, Christy K.
and Dommer, Abigail C.
and Harbison, Aoife M.
and Fogarty, Carl A.
and Barros, Emilia P.
and Taylor, Bryn C.
and McLellan, Jason S.
and Fadda, Elisa
and Amaro, Rommie E.},
title={Beyond Shielding: The Roles of Glycans in the SARS-CoV-2 Spike Protein},
journal={ACS Central Science},
year={2020},
month={Oct},
day={28},
publisher={American Chemical Society},
volume={6},
number={10},
pages={1722-1734},
issn={2374-7943},
}

@Article{Posani2025,
author={Posani, Elisa
and Jano{\v{s}}, Pavel
and Haack, Daniel
and Toor, Navtej
and Bonomi, Massimiliano
and Magistrato, Alessandra
and Bussi, Giovanni},
title={Ensemble refinement of mismodeled cryo-EM RNA structures using all-atom simulations},
journal={Nature Communications},
year={2025},
month={May},
day={16},
volume={16},
number={1},
pages={4549},
}

@article{Shaw2013,
title = {Molecular determinants of drug–receptor binding kinetics},
journal = {Drug Discovery Today},
volume = {18},
number = {13},
pages = {667-673},
year = {2013},
issn = {1359-6446},
author = {Albert C. Pan and David W. Borhani and Ron O. Dror and David E. Shaw},
}

@article{Shaw2011,
author = {Kresten Lindorff-Larsen  and Stefano Piana  and Ron O. Dror  and David E. Shaw },
title = {How Fast-Folding Proteins Fold},
journal = {Science},
volume = {334},
number = {6055},
pages = {517-520},
year = {2011},
}

@article{Dror2012,
   author = "Dror, Ron O. and Dirks, Robert M. and Grossman, J.P. and Xu, Huafeng and Shaw, David E.",
   title = "Biomolecular Simulation: A Computational Microscope for Molecular Biology", 
   journal= "Annual Review of Biophysics",
   year = "2012",
   volume = "41",
   number = "Volume 41, 2012",
   pages = "429-452",
   publisher = "Annual Reviews",
   issn = "1936-1238",  
}

@Article{Unke2021,
author={Unke, Oliver T.
and Chmiela, Stefan
and Sauceda, Huziel E.
and Gastegger, Michael
and Poltavsky, Igor
and Sch{\"u}tt, Kristof T.
and Tkatchenko, Alexandre
and M{\"u}ller, Klaus-Robert},
title={Machine Learning Force Fields},
journal={Chemical Reviews},
year={2021},
month={Aug},
day={25},
publisher={American Chemical Society},
volume={121},
number={16},
pages={10142-10186},

}

@article{Rahman1964,
  title = {Correlations in the Motion of Atoms in Liquid Argon},
  author = {Rahman, A.},
  journal = {Phys. Rev.},
  volume = {136},
  issue = {2A},
  pages = {A405--A411},
  numpages = {0},
  year = {1964},
  month = {Oct},
  publisher = {American Physical Society}
}

@Article{Li2020,
author ="Li, Zhiheng and Wellawatte, Geemi P. and Chakraborty, Maghesree and Gandhi, Heta A. and Xu, Chenliang and White, Andrew D.",
title  ="Graph neural network based coarse-grained mapping prediction",
journal  ="Chem. Sci.",
year  ="2020",
volume  ="11",
issue  ="35",
pages  ="9524-9531",
publisher  ="The Royal Society of Chemistry",
}

@article{Webb2019,
author = {Webb, Michael A. and Delannoy, Jean-Yves and de Pablo, Juan J.},
title = {Graph-Based Approach to Systematic Molecular Coarse-Graining},
journal = {Journal of Chemical Theory and Computation},
volume = {15},
number = {2},
pages = {1199-1208},
year = {2019},
}

@article{Machado2016,
    author = {Machado, Matías R. and Pantano, Sergio},
    title = {SIRAH tools: mapping, backmapping and visualization of coarse-grained models},
    journal = {Bioinformatics},
    volume = {32},
    number = {10},
    pages = {1568-1570},
    year = {2016},
    month = {01},
}

@article{Rudzinski2014,
author = {Rudzinski, Joseph F. and Noid, William G.},
title = {Investigation of Coarse-Grained Mappings via an Iterative Generalized Yvon–Born–Green Method},
journal = {The Journal of Physical Chemistry B},
volume = {118},
number = {28},
pages = {8295-8312},
year = {2014},
}

@article{Voth2018,
title = {Advances in coarse-grained modeling of macromolecular complexes},
journal = {Current Opinion in Structural Biology},
volume = {52},
pages = {119-126},
year = {2018},
issn = {0959-440X},
author = {Alexander J Pak and Gregory A Voth},
}

@article{Zhong2025,
author = {Zhong, Zhixuan and Xu, Lifeng and Jiang, Jian},
title = {A Neural-Network-Based Mapping and Optimization Framework for High-Precision Coarse-Grained Simulation},
journal = {Journal of Chemical Theory and Computation},
volume = {21},
number = {2},
pages = {859-870},
year = {2025},
}

@article{Voth2021,
author = {Jarin, Zack and Newhouse, James and Voth, Gregory A.},
title = {Coarse-Grained Force Fields from the Perspective of Statistical Mechanics: Better Understanding of the Origins of a MARTINI Hangover},
journal = {Journal of Chemical Theory and Computation},
volume = {17},
number = {2},
pages = {1170-1180},
year = {2021},
}

@article{Weinan2018,
    author = {Zhang, Linfeng and Han, Jiequn and Wang, Han and Car, Roberto and E, Weinan},
    title = {DeePCG: Constructing coarse-grained models via deep neural networks},
    journal = {The Journal of Chemical Physics},
    volume = {149},
    number = {3},
    pages = {034101},
    year = {2018},
    month = {07},
}

@article{Kremer2015,
author = {Bereau, Tristan and Kremer, Kurt},
title = {Automated Parametrization of the Coarse-Grained Martini Force Field for Small Organic Molecules},
journal = {Journal of Chemical Theory and Computation},
volume = {11},
number = {6},
pages = {2783-2791},
year = {2015},
}

@article{Tang2023,
author = {Mahajan, Subhamoy and Tang, Tian},
title = {Automated Parameterization of Coarse-Grained Polyethylenimine under a Martini Framework},
journal = {Journal of Chemical Information and Modeling},
volume = {63},
number = {14},
pages = {4328-4341},
year = {2023},
}

@Article{Wang2019,
author={Wang, Wujie
and G{\'o}mez-Bombarelli, Rafael},
title={Coarse-graining auto-encoders for molecular dynamics},
journal={npj Computational Materials},
year={2019},
month={Dec},
day={18},
volume={5},
number={1},
pages={125},
}

@article{Doniach1999,
title = {Protein dynamics simulations from nanoseconds to microseconds},
journal = {Current Opinion in Structural Biology},
volume = {9},
number = {2},
pages = {157-163},
year = {1999},
issn = {0959-440X},
author = {Sebastian Doniach and Peter Eastman},
}

@Article{Freddolino2008,
author={Freddolino, Lydia
and Liu, Feng
and Gruebele, Martin
and Schulten, Klaus},
title={Ten-Microsecond Molecular Dynamics Simulation of a Fast-Folding WW Domain},
journal={Biophysical Journal},
year={2008},
month={May},
day={15},
publisher={Elsevier},
volume={94},
number={10},
pages={L75-L77},
issn={0006-3495},
}

@incollection{Feig2019,
    author = {Sugita, Yuji and Feig, Michael},
    isbn = {978-1-78801-217-1},
    title = {All-atom Molecular Dynamics Simulation of Proteins in Crowded Environments},
    booktitle = {In-cell NMR Spectroscopy: From Molecular Sciences to Cell Biology},
    publisher = {The Royal Society of Chemistry},
    year = {2019},
    month = {12},
}

@article{Gunsteren2004,
author = {Oostenbrink, Chris and Villa, Alessandra and Mark, Alan E. and Van Gunsteren, Wilfred F.},
title = {A biomolecular force field based on the free enthalpy of hydration and solvation: The GROMOS force-field parameter sets 53A5 and 53A6},
journal = {Journal of Computational Chemistry},
volume = {25},
number = {13},
pages = {1656-1676},
year = {2004},
}

@Article{Jorgensen1988,
author={Jorgensen, William L.
and Tirado-Rives, Julian},
title={The OPLS [optimized potentials for liquid simulations] potential functions for proteins, energy minimizations for crystals of cyclic peptides and crambin},
journal={Journal of the American Chemical Society},
year={1988},
month={Mar},
day={01},
publisher={American Chemical Society},
volume={110},
number={6},
pages={1657-1666},
issn={0002-7863},
}

@Article{Souza2020,
author={Souza, Paulo C. T.
and Thallmair, Sebastian
and Conflitti, Paolo
and Ram{\'i}rez-Palacios, Carlos
and Alessandri, Riccardo
and Raniolo, Stefano
and Limongelli, Vittorio
and Marrink, Siewert J.},
title={Protein--ligand binding with the coarse-grained Martini model},
journal={Nature Communications},
year={2020},
month={Jul},
day={24},
volume={11},
number={1},
pages={3714},
}

@Article{Kjolbye2022,
author={Kj{\o}lbye, Lisbeth R.
and Pereira, Gilberto P.
and Bartocci, Alessio
and Pannuzzo, Martina
and Albani, Simone
and Marchetto, Alessandro
and Jim{\'e}nez-Garc{\'i}a, Brian
and Martin, Juliette
and Rossetti, Giulia
and Cecchini, Marco
and Wu, Sangwook
and Monticelli, Luca
and Souza, Paulo C. T.},
title={Towards design of drugs and delivery systems with the Martini coarse-grained model},
journal={QRB Discovery},
year={2022},
edition={2022/10/12},
publisher={Cambridge University Press},
volume={3},
pages={e19},
}

@misc{Kaggle,
  author       = {matthewmasters},
  title        = {LogP of Chemical Structures},
  year         = {2025},
  howpublished = {\url{https://www.kaggle.com/datasets/matthewmasters/chemical-structure-and-logp}},
  note         = {Kaggle dataset; accessed 2025-09-06},
}

@article{2Da,
author = {Wu, Kedi and Zhao, Zhixiong and Wang, Renxiao and Wei, Guo-Wei},
title = {TopP–S: Persistent homology-based multi-task deep neural networks for simultaneous predictions of partition coefficient and aqueous solubility},
journal = {Journal of Computational Chemistry},
volume = {39},
number = {20},
pages = {1444-1454},
doi = {https://doi.org/10.1002/jcc.25213},
url = {https://onlinelibrary.wiley.com/doi/abs/10.1002/jcc.25213},
eprint = {https://onlinelibrary.wiley.com/doi/pdf/10.1002/jcc.25213},
year = {2018}
}

@Article{2Db,
author={Chen, Dong
and Gao, Kaifu
and Nguyen, Duc Duy
and Chen, Xin
and Jiang, Yi
and Wei, Guo-Wei
and Pan, Feng},
title={Algebraic graph-assisted bidirectional transformers for molecular property prediction},
journal={Nature Communications},
year={2021},
month={Jun},
day={10},
volume={12},
number={1},
pages={3521},
}

@article{Szczuka2025,
author = {Szczuka, Magdalena and Pereira, Gilberto P. and Walter, Luis J. and Gueroult, Marc and Poulain, Pierre and Bereau, Tristan and Souza, Paulo C. T. and Chavent, Matthieu},
title = {Fast Parametrization of Martini3 Models for Fragments and Small Molecules},
journal = {Journal of Chemical Theory and Computation},
volume = {22},
number = {1},
pages = {610-623},
year = {2026},
}

@misc{TPCN2024,
  author       = {Kong Lab},
  title        = {{TPCN}: Terpenoids Content Database (Version 1.0)},
  year         = {2024},
  institution  = {National Key Laboratory of Agricultural Microbiology, College of Information Science and Technology, Huazhong Agricultural University, Wuhan, Hubei, PR China},
  howpublished = {\url{http://www.tpcn.pro}},
}

@misc{RDKit,
  author = {Greg Landrum},
  title = {RDKit: Open-source cheminformatics},
  year = {2006--},
  howpublished = {\url{http://www.rdkit.org}}
}

@article{Kirkwood,
    author = {Kirkwood, John G.},
    title = {Statistical Mechanics of Fluid Mixtures},
    journal = {The Journal of Chemical Physics},
    volume = {3},
    number = {5},
    pages = {300-313},
    year = {1935},
    month = {05},
}

@article{weininger1988smiles,
  title={SMILES, a chemical language and information system. 1. Introduction to methodology and encoding rules},
  author={Weininger, David},
  journal={Journal of chemical information and computer sciences},
  volume={28},
  number={1},
  pages={31--36},
  year={1988},
  publisher={ACS Publications}
}

@article{bejagam2018machine,
  title={Machine-learned coarse-grained models},
  author={Bejagam, Karteek K and Singh, Samrendra and An, Yaxin and Deshmukh, Sanket A},
  journal={The journal of physical chemistry letters},
  volume={9},
  number={16},
  pages={4667--4672},
  year={2018},
  publisher={ACS Publications}
}

@article{an2020machine,
  title={Machine learning approach for accurate backmapping of coarse-grained models to all-atom models},
  author={An, Yaxin and Deshmukh, Sanket A},
  journal={Chemical communications},
  volume={56},
  number={65},
  pages={9312--9315},
  year={2020},
  publisher={Royal Society of Chemistry}
}

@article{an2018development,
  title={Development of new transferable coarse-grained models of hydrocarbons},
  author={An, Yaxin and Bejagam, Karteek K and Deshmukh, Sanket A},
  journal={The Journal of Physical Chemistry B},
  volume={122},
  number={28},
  pages={7143--7153},
  year={2018},
  publisher={ACS Publications}
}

@article{bartender,
author = {Pereira, Gilberto P. and Alessandri, Riccardo and Domínguez, Mois{\'e}s and Araya-Osorio, Rocío and Gr{\"u}newald, Linus and Borges-Araújo, Luís and Wu, Sangwook and Marrink, Siewert J. and Souza, Paulo C. T. and Mera-Adasme, Raul},
title = {Bartender: Martini 3 Bonded Terms via Quantum Mechanics-Based Molecular Dynamics},
journal = {Journal of Chemical Theory and Computation},
volume = {20},
number = {13},
pages = {5763-5773},
year = {2024},
}

@article{pycgtool,
author = {Graham, James A. and Essex, Jonathan W. and Khalid, Syma},
title = {PyCGTOOL: Automated Generation of Coarse-Grained Molecular Dynamics Models from Atomistic Trajectories},
journal = {Journal of Chemical Information and Modeling},
volume = {57},
number = {4},
pages = {650-656},
year = {2017},
}

@article{Vanikka,
author = {Vainikka, Petteri and Marrink, Siewert J.},
title = {Martini 3 Coarse-Grained Model for Second-Generation Unidirectional Molecular Motors and Switches},
journal = {Journal of Chemical Theory and Computation},
volume = {19},
number = {2},
pages = {596-604},
year = {2023},
}

@article{genff,
author = {Spicher, Sebastian and Grimme, Stefan},
title = {Robust Atomistic Modeling of Materials, Organometallic, and Biochemical Systems},
journal = {Angewandte Chemie International Edition},
volume = {59},
number = {36},
pages = {15665-15673},
year = {2020}
}

@article{xTB,
author = {Bannwarth, Christoph and Caldeweyher, Eike and Ehlert, Sebastian and Hansen, Andreas and Pracht, Philipp and Seibert, Jakob and Spicher, Sebastian and Grimme, Stefan},
title = {Extended tight-binding quantum chemistry methods},
journal = {WIREs Computational Molecular Science},
volume = {11},
number = {2},
pages = {e1493},
year = {2021}
}

@Article{Bento2020,
author={Bento, A. Patr{\'i}cia
and Hersey, Anne
and F{\'e}lix, Eloy
and Landrum, Greg
and Gaulton, Anna
and Atkinson, Francis
and Bellis, Louisa J.
and De Veij, Marleen
and Leach, Andrew R.},
title={An open source chemical structure curation pipeline using RDKit},
journal={Journal of Cheminformatics},
year={2020},
month={Sep},
day={01},
volume={12},
number={1},
pages={51},
}

@article{ETKDG,
author = {Wang, Shuzhe and Witek, Jagna and Landrum, Gregory A. and Riniker, Sereina},
title = {Improving Conformer Generation for Small Rings and Macrocycles Based on Distance Geometry and Experimental Torsional-Angle Preferences},
journal = {Journal of Chemical Information and Modeling},
volume = {60},
number = {4},
pages = {2044-2058},
year = {2020},
}

@article{Grunewald_2025,
author = {Gr{\"u}newald, Fabian and Seute, Leif and Alessandri, Riccardo and K{\"o}nig, Melanie and Kroon, Peter C.},
title = {CGsmiles: A Versatile Line Notation for Molecular Representations across Multiple Resolutions},
journal = {Journal of Chemical Information and Modeling},
volume = {65},
number = {7},
pages = {3405-3419},
year = {2025},
}

@article{exp1,
    author = {Sangster, James},
    title = {Octanol‐Water Partition Coefficients of Simple Organic Compounds},
    journal = {Journal of Physical and Chemical Reference Data},
    volume = {18},
    number = {3},
    pages = {1111-1229},
    year = {1989},
    month = {07},
}

@book{exp2,
  title={Exploring QSAR: hydrophobic, electronic, and steric constants},
  author={Hansch, Corwin and Leo, Albert and Hoekman, David and others},
  volume={2},
  year={1995},
  publisher={American Chemical Society Washington, DC}
}

@article{exp3,
author = {Abraham, Michael H. and Chadha, Harpreet S. and Whiting, Gary S. and Mitchell, Robert C.},
title = {Hydrogen bonding. 32. An analysis of water-octanol and water-alkane partitioning and the $\delta$log p parameter of seiler},
journal = {Journal of Pharmaceutical Sciences},
volume = {83},
number = {8},
pages = {1085-1100},
year = {1994}
}

@article{exp4,
author = {Klamt, Andreas and Jonas, Volker and B{\"u}rger, Thorsten and Lohrenz, John C. W.},
title = {Refinement and Parametrization of COSMO-RS},
journal = {The Journal of Physical Chemistry A},
volume = {102},
number = {26},
pages = {5074-5085},
year = {1998},
}

@article{exp5,
author = {Abraham, Michael H. and Platts, James A. and Hersey, Anne and Leo, Albert J. and Taft, Robert W.},
title = {Correlation and estimation of gas–chloroform and water–chloroform partition coefficients by a linear free energy relationship method},
journal = {Journal of Pharmaceutical Sciences},
volume = {88},
number = {7},
pages = {670-679},
year = {1999}
}

@article{exp6,
author = {Natesan, Senthil and Wang, Zhanbin and Lukacova, Viera and Peng, Ming and Subramaniam, Rajesh and Lynch, Sandra and Balaz, Stefan},
title = {Structural Determinants of Drug Partitioning in n-Hexadecane/Water System},
journal = {Journal of Chemical Information and Modeling},
volume = {53},
number = {6},
pages = {1424-1435},
year = {2013},
}

@incollection{Souza_book,
    author = {Alessandri, Riccardo and Thallmair, Sebastian and Herrero, Cristina Gil and Mera-Adasme, Raúl and Marrink, Siewert J. and Souza, Paulo C. T.},
    isbn = {978-0-7354-2524-8},
    title = {A Practical Introduction to Martini 3 and its Application to Protein-Ligand Binding Simulations},
    booktitle = {A Practical Guide to Recent Advances in Multiscale Modeling and Simulation of Biomolecules},
    publisher = {AIP Publishing LLC},
    year={2023}
}

@misc{martini3_tutorials,
  author       = {{Martini}},
  title        = {Martini 3 Tutorials},
  year         = {2025},
  url          = {https://cgmartini.nl/docs/tutorials/Martini3/tutorials.html},
}

@Article{Tosco2014,
author={Tosco, Paolo
and Stiefl, Nikolaus
and Landrum, Gregory},
title={Bringing the MMFF force field to the RDKit: implementation and validation},
journal={Journal of Cheminformatics},
year={2014},
month={Jul},
day={12},
volume={6},
number={1},
pages={37},
}

@Article{Pedersen2025,
author={Pedersen, Kasper B.
and Ing{\'o}lfsson, Helgi I.
and Ramirez-Echemendia, Daniel P.
and Borges-Ara{\'u}jo, Lu{\'i}s
and Andreasen, Mikkel D.
and Empereur-mot, Charly
and Melcr, Josef
and Ozturk, Tugba N.
and Bennett, W. F. Drew
and Kj{\o}lbye, Lisbeth R.
and Brasnett, Christopher
and Corradi, Valentina
and Khan, Hanif M.
and Cino, Elio A.
and Crowley, Jackson
and Kim, Hyuntae
and F{\'a}bi{\'a}n, Bal{\'a}zs
and Borges-Ara{\'u}jo, Ana C.
and Pavan, Giovanni M.
and Launay, Guillaume
and Lolicato, Fabio
and Wassenaar, Tsjerk A.
and Melo, Manuel N.
and Thallmair, Sebastian
and Carpenter, Timothy S.
and Monticelli, Luca
and Tieleman, D. Peter
and Schi{\o}tt, Birgit
and Souza, Paulo C. T.
and Marrink, Siewert J.},
title={The Martini 3 Lipidome: Expanded and Refined Parameters Improve Lipid Phase Behavior},
journal={ACS Central Science},
year={2025},
month={Sep},
day={24},
publisher={American Chemical Society},
volume={11},
number={9},
pages={1598-1610},
}

@Article{Vazquez2020,
author ="Vazquez-Salazar, Luis Itza and Selle, Michele and de Vries, Alex H. and Marrink, Siewert J. and Souza, Paulo C. T.",
title  ="Martini coarse-grained models of imidazolium-based ionic liquids: from nanostructural organization to liquid–liquid extraction",
journal  ="Green Chem.",
year  ="2020",
volume  ="22",
issue  ="21",
pages  ="7376-7386",
publisher  ="The Royal Society of Chemistry",
}

@Article{polyply,
author={Gr{\"u}newald, Fabian
and Alessandri, Riccardo
and Kroon, Peter C.
and Monticelli, Luca
and Souza, Paulo C. T.
and Marrink, Siewert J.},
title={Polyply; a python suite for facilitating simulations of macromolecules and nanomaterials},
journal={Nature Communications},
year={2022},
month={Jan},
day={10},
volume={13},
number={1},
pages={68},
}

@article{PyLipID,
author = {Song, Wanling and Corey, Robin A. and Ansell, T. Bertie and Cassidy, C. Keith and Horrell, Michael R. and Duncan, Anna L. and Stansfeld, Phillip J. and Sansom, Mark S. P.},
title = {PyLipID: A Python Package for Analysis of Protein–Lipid Interactions from Molecular Dynamics Simulations},
journal = {Journal of Chemical Theory and Computation},
volume = {18},
number = {2},
pages = {1188-1201},
year = {2022},
}

\end{document}